\documentstyle[12pt]{article}

\catcode`\@=11
\def\numberbysection{\@addtoreset{equation}{section}
	\def\theequation{\thesection.\arabic{equation}}}
\def\beq{\begin{equation}}
\def\eeq{\end{equation}}
\def\barr{\begin{eqnarray}}
\def\earr{\end{eqnarray}}
\def\winf{W_{1+\infty}\ }
\def\u1{\widehat{U(1)}}
\def\rep{ representation }
\def\reps{ representations }
\def\scr{\scriptstyle}

\numberbysection
\begin{document}
\begin{titlepage}
\begin{center}
\hfill UGVA-DPT 1995/01-879 \quad DFTT 09/95 \\
\vskip .3 in
{\large \bf Stable Hierarchical Quantum Hall Fluids as \\
 $\winf$ Minimal Models   }
\vskip 0.2in
Andrea CAPPELLI \\
{\em I.N.F.N. and Dipartimento di Fisica, Largo E. Fermi 2,
 I-50125 Firenze, Italy}
\\
\vskip 0.1in
Carlo~A.~TRUGENBERGER\footnote{Supported by a Profil 2 fellowship of
the Swiss National Science Foundation.}  \\
{\em D\'epartement de Physique Th\'eorique, Univ. de Gen\`eve,
24 quai E. Ansermet, CH-1211, Gen\`eve 4, Switzerland}
\\
\vskip 0.1in
Guillermo~R.~ZEMBA \\%[.2in]
{\em I.N.F.N. and Dipartimento di Fisica Teorica,
	  Via P. Giuria 1, I-10125 Torino, Italy}
\end{center}
\vskip .1 in
\begin{abstract}
In this paper, we pursue our analysis of the $\winf$ symmetry of the low-energy
edge excitations of incompressible quantum Hall fluids.
These excitations are described by $(1+1)$-dimensional effective field
theories, which are built by representations of the $\winf$ algebra.
Generic $\winf$ theories predict many more fluids than the few,
stable ones found in experiments.
Here we identify a particular class of $\winf$ theories, the
minimal models, which are made of degenerate representations only -
a familiar construction in conformal field theory.
The $\winf$ minimal models exist for specific values of the fractional
conductivity, which nicely fit the experimental data and match
the results of the Jain hierarchy of quantum Hall fluids.
We thus obtain a new hierarchical construction, which is based uniquely
on the concept of quantum incompressible fluid and is independent of
Jain's approach and hypotheses.
Furthermore, a surprising non-Abelian structure is found in
the $\winf$ minimal models: they possess neutral quark-like excitations
with $SU(m)$ quantum numbers and non-Abelian fractional statistics.
The physical electron is made of anyon and quark excitations.
We discuss some properties of these neutral excitations which could be seen
in experiments and numerical simulations.
\end{abstract}
\vfill
January 1995\hfill\\
\end{titlepage}
\pagenumbering{arabic}
%
%-1-------------

\section{Introduction}

In the past few years, the fractional quantum Hall effect \cite{prange}
has become an exciting arena for new physics.
Very precise measurements have stimulated
the theoretical physicists beyond the borders of the solid-state community.
Many ideas, and sophisticated mathematical tools, have been proposed
to describe the non-trivial quantum many-body ground states revealed by the
experiments.
More recently, new experiments have remarkably confirmed some of the
new key ideas. For example, the new concepts of {\it edge excitation }
and {\it composite fermion} have acquired the status of real quantum
many-body states.

\begin{figure}
\unitlength=0.8pt
\begin{picture}(500.00,440.00)(-10.00,0.00)
\put(0.00,500.00){\line(1,0){500.00}}
\put(0.00,80.00){\line(1,0){500.00}}
\put(250.00,500.00){\line(0,-1){100.00}}
\put(125.00,500.00){\line(0,-1){100.00}}
\put(83.00,500.00){\line(0,-1){100.00}}
\put(250.00,80.00){\line(0,1){40.00}}
\put(125.00,80.00){\line(0,1){100.00}}
\put(83.00,80.00){\line(0,1){100.00}}
\put(166.00,460.00){\makebox(0,0)[cc]{$\ \ \ \bullet {\bf 1\over 3}$}}
\put(333.00,460.00){\makebox(0,0)[cc]{$\ \ \ \bullet {\bf 2\over 3}$}}
\put(100.00,420.00){\makebox(0,0)[cc]{$\ \ \ \bullet {\bf 1\over 5}$}}
\put(200.00,420.00){\makebox(0,0)[cc]{$\ \ \ \bullet {\bf 2\over 5}$}}
\put(300.00,420.00){\makebox(0,0)[cc]{$\ \ \ \bullet {\bf 3\over 5}$}}
\put(400.00,420.00){\makebox(0,0)[cc]{$\ \ \ \bullet {\it4\over 5}$}}
\put(71.00,380.00){\makebox(0,0)[cc]{$\ \ \ \circ    {\bf 1\over 7}$}}
\put(143.00,380.00){\makebox(0,0)[cc]{$\ \ \ \bullet {\bf 2\over 7}$}}
\put(214.00,380.00){\makebox(0,0)[cc]{$\ \ \ \bullet {\bf 3\over 7}$}}
\put(286.00,380.00){\makebox(0,0)[cc]{$\ \ \ \bullet {\bf 4\over 7}$}}
\put(357.00,380.00){\makebox(0,0)[cc]{$\ \ \ \bullet {\it 5\over 7}$}}
\put(111.00,340.00){\makebox(0,0)[cc]{$\ \ \ \bullet {\bf 2\over 9}$}}
\put(222.00,340.00){\makebox(0,0)[cc]{$\ \ \ \bullet {\bf 4\over 9}$}}
\put(278.00,340.00){\makebox(0,0)[cc]{$\ \ \ \bullet {\bf 5\over 9}$}}
\put(333.00,340.00){\makebox(0,0)[cc]{$\ \ \ \bullet {\it 6\over 9}$}}
\put(91.00,300.00){\makebox(0,0)[cc]{$\ \ \ \circ    {\bf 2\over 11}$}}
\put(136.00,300.00){\makebox(0,0)[cc]{$\ \ \ \bullet {\bf 3\over 11}$}}
\put(182.00,300.00){\makebox(0,0)[cc]{$\ \ \ \cdot   {\it 4\over 11}$}}
\put(227.00,300.00){\makebox(0,0)[cc]{$\ \ \ \bullet {\bf 5\over 11}$}}
\put(273.00,300.00){\makebox(0,0)[cc]{$\ \ \ \bullet {\bf 6\over 11}$}}
\put(318.00,300.00){\makebox(0,0)[cc]{$\ \ \ \circ   {\it7\over 11}$}}
\put(367.00,300.00){\makebox(0,0)[cc]{$\ \ \ \bullet {\it8\over 11}$}}
\put(115.00,260.00){\makebox(0,0)[cc]{$\ \ \ \circ   {\bf 3\over 13}$}}
\put(154.00,260.00){\makebox(0,0)[cc]{$\ \ \ \cdot   {\it4\over 13}$}}
\put(231.00,260.00){\makebox(0,0)[cc]{$\ \ \ \bullet {\bf 6\over 13}$}}
\put(269.00,260.00){\makebox(0,0)[cc]{$\ \ \ \bullet {\bf 7\over 13}$}}
\put(308.00,260.00){\makebox(0,0)[cc]{$\ \ \ \bullet {\it8\over 13}$}}
\put(346.00,260.00){\makebox(0,0)[cc]{$\ \ \ \circ   {\it9\over 13}$}}
\put(133.00,220.00){\makebox(0,0)[cc]{$\ \ \ \circ   {\bf 4\over 15}$}}
\put(233.00,220.00){\makebox(0,0)[cc]{$\ \ \ \circ   {\bf 7\over 15}$}}
\put(267.00,220.00){\makebox(0,0)[cc]{$\ \ \ \circ   {\bf 8\over 15}$}}
\put(300.00,220.00){\makebox(0,0)[cc]{$\ \ \ \bullet {\it9\over 15}$}}
\put(333.00,220.00){\makebox(0,0)[cc]{$\ \ \ \bullet {\it 10\over 15}$}}
\put(235.00,180.00){\makebox(0,0)[cc]{$\ \ \ \circ   {\bf 8\over 17}$}}
\put(265.00,180.00){\makebox(0,0)[cc]{$\ \ \ \circ   {\bf 9\over 17}$}}
\put(294.00,180.00){\makebox(0,0)[cc]{$\ \ \ \cdot   {\it10\over 17}$}}
\put(237.00,140.00){\makebox(0,0)[cc]{$\ \ \ \cdot   {\bf 9\over 19}$}}
\put(83.00,60.00){\makebox(0,0)[cc]{${\bf 1\over 6}$}}
\put(125.00,60.00){\makebox(0,0)[cc]{${\bf 1\over 4}$}}
\put(250.00,60.00){\makebox(0,0)[cc]{${\bf 1\over 2}$}}
\put(0.00,60.00){\makebox(0,0)[cc]{${\bf 0}$}}
\put(500.00,60.00){\makebox(0,0)[rc]{$\sigma_H \qquad {\bf 1}$}}
\end{picture}
\caption{
Experimentally observed plateaus in the range $0<\sigma_H<1$ : their Hall
conductivity $\sigma_H=(e^2/h)\nu$ is displayed in units of $(e^2/h)$.
The points denote stability:
$\ (\bullet)\ $ very stable, $\ (\circ)\ $ stable, and
$\ (\cdot)\ $ less stable plateaus. Theoretically understood plateaus are in
{\bf bold}, unexplained ones are in {\it italic}.
Observed cases of coexisting fluids are displayed
as $\nu=2/3,6/9,10/15$,  $\ \nu=3/5,9/15$ and $\nu=5/7, 15/21$ (but
$15/21$ is not displayed). (Adapted from ref. [26])}
\end{figure}
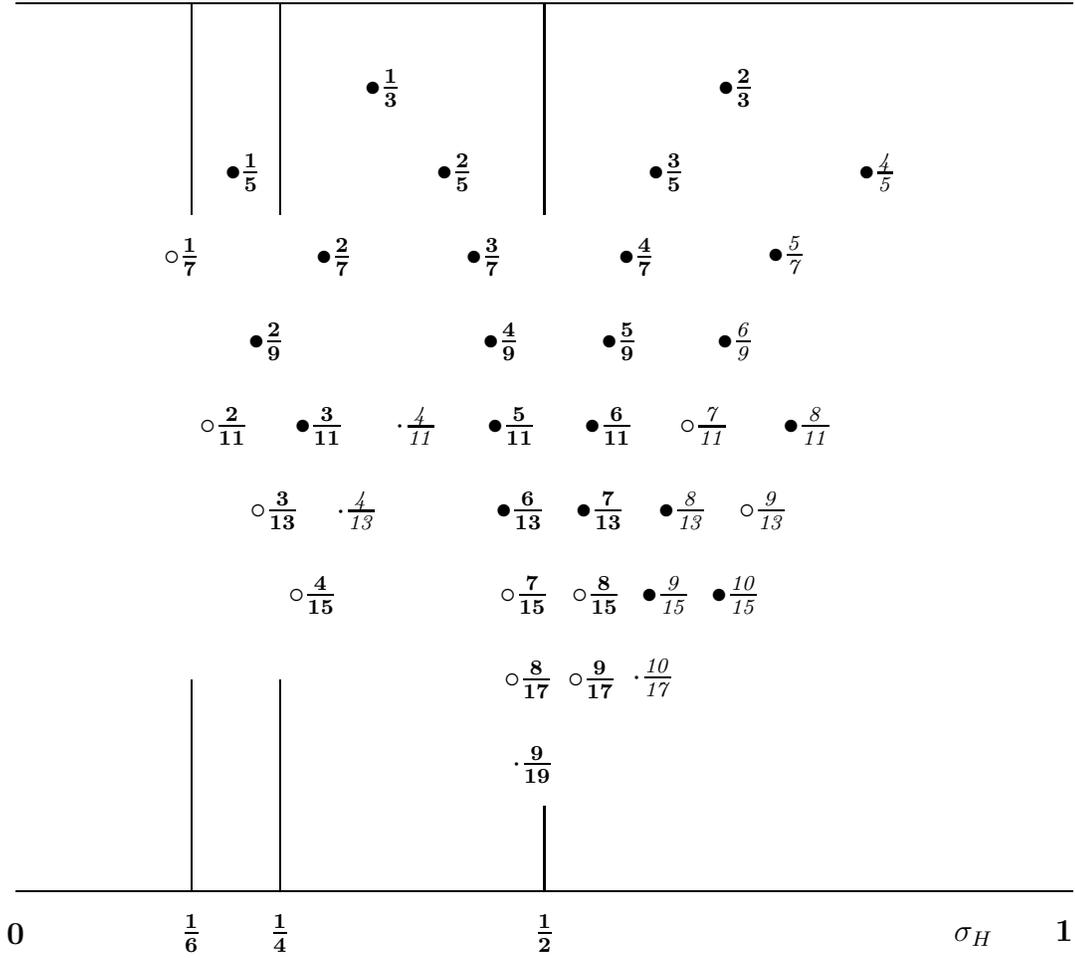

\bigskip
\noindent{\bf Hierarchical trial wave functions}

There are many aspects of the quantum Hall effect, which reflect the
dynamics at different scales of energy and ranges of the control
parameters. The problem of our concern here is the description of the
stable ground states of the electrons, corresponding to the plateaus
of the Hall conductivity, and their spectrum of low-lying excitations.
These are relevant for the precise conduction experiments.

Laughlin \cite{laugh} introduced the key concept of the {\it quantum
incompressible fluid} of electrons, which is a ground state with
uniform density and a gap for density fluctuations.
This ground state, as well as its low-lying excitations, was originally
described by trial wave functions for the Hall conductivities
$\ \sigma_{xy}=(e^2/h) \nu\ $, where $\ \nu=1,1/3,1/5,1/7,\dots$
are the filling fractions.
Afterwards, a hierarchical generalization of these trial wave functions
was introduced by Haldane and Halperin \cite{hald}, in order to
describe other observed filling fractions.
Therefore, by the {\it hierarchy} problem we usually mean the classification
of stable ground states (and their excitations) corresponding
to all observed plateaus.
Naturally, the stability is related to the order of iteration
of the hierarchical construction, starting from the integer fillings,
then the Laughlin fillings and so forth.

The Haldane-Halperin hierarchy is not completely satisfactory, because
it produces ground states for too many filling fractions, already
at low order of iteration.
On the contrary, the experiments show only some stable ground states
(see fig. 1).
Although numerical experiments show that the hierarchical wave
functions are rather accurate, their construction lacks a good control
of stability.

Another hierarchical construction of wave functions, which match most of
the experimental plateaus to lowest order of the hierarchy, has been
proposed by Jain \cite{jain}.
Jain abstracted from Laughlin's work the concept of {\it composite
fermion}, a local bound state of the electron and an even number of
flux quanta.
Due to yet unknown dynamical reasons, the composite fermions are
stable quasi-particles, which interact weakly among themselves.
Moreover, the strongly-interacting electrons at fractional filling can be
mapped into composite fermions at effective integer filling.
Therefore, the stability of the observed ground states
with fractional filling can be related with the stability of completely
filled Landau levels.
The composite fermion picture was successfully applied \cite{hlr} to the
independent dynamics of the compressible fluid at $\nu=1/2$.
This strongly-interacting, gapless ground state can be described
as a Fermi liquid of composite fermions with vanishing effective magnetic
field.
Experiments \cite{nu1/2} have confirmed this theory by observing the free
motion of the composite fermions.

\bigskip
\noindent{\bf Theories of edge excitations}

In open domains of typical size $R$, like an annulus, the Laughlin
incompressible fluid possesses ``gapless'' excitations of energy $O(1/R)$
located at the boundary \cite{halp}\cite{stone}\cite{wen}.
These are shape deformations of the fluid edge at
constant density (bulk density modulations are suppressed by the gap).
The dynamics of the quantum incompressible fluid for energies below the
gap is completely described by a $(1+1)$-dimensional
field theory of the edge excitations, which is defined on the two edge
circles.
The propagation of low-energy waves along the edge, with
definite chirality (either clockwise or anti-clockwise), has been clearly
estabilished by an experiment with precise time resolution \cite{tdom}.
Moreover, their $O(1/R)$ energy spectrum has been measured by radio-frequency
resonance \cite{wasser}. Finally, the spectrum of edge excitations and
their $(1+1)$-dimensional interaction has been
confirmed by the experiment of resonant tunnelling through a point
contact \cite{milli}. Two edges of the electron fluid are
pinched by applying a localized voltage, such that edge excitations can
interact; the resonant conductance is a computable universal scaling
function \cite{tunn}, which fits the experimental data.

The theories of edge excitations are effective field theories of the
quantum incompressible fluids, whose
low-energy, long-distance dynamics is universal in the sense of the
renormalization group \cite{frad}\cite{cdtz2}.
The long-distance {\it scaling} limit defining the edge theories can be
explicitly performed at integer fillings \cite{stone}\cite{cdtz1}\cite{sak}.
These $(1+1)$-dimensional field theories are conformal field theories, which
can be solved exactly \cite{bpz}. Although for a limited energy range,
they yield exact kinematical properties of the incompressible fluid, which are
sufficient to describe the precise conduction experiments.

After the original works of Halperin \cite{halp} and Stone \cite{stone},
a general theory of edge excitations, corresponding to the hierarchical
constructions of wave-functions, has been formulated
\cite{jtrans}\cite{wen}\cite{kmat}.
This is the $(1+1)$-dimensional theory of the chiral boson \cite{flo}.
An equivalent description is given by
Abelian Chern-Simons theories on $(2+1)$-dimensional open domains \cite{kmat}.
The edge excitations of the Laughlin fluid are described by a one-component
chiral boson, while the hierarchical fluids require many components.
Every boson describes an independent edge current, and thus the
incompressible fluids have generically a composite edge structure.
Each current gives rise to the Abelian current algebra in $(1+1)$-dimensions,
denoted by $\widehat{U(1)}$, which implies the Virasoro algebra with
central charge $c=1$ \cite{bpz}.
These are well-known examples of the infinite-dimensional
symmetries in $(1+1)$ dimensions.

The $m$-edge theory has $\u1^{\otimes m}$ symmetry, $c=m$, and is parametrized
by an integer, symmetric $(m\times m)$ matrix, with odd diagonal elements,
the $K$ matrix \cite{kmat}.
This determines the Hall conductivity and the fractional charge, spin
and statistics \cite{wilc} of the edge excitations, which correspond to
the anyon quasi-particles of the incompressible fluid \cite{laugh}.
The Haldane-Halperin and the Jain hierarchies of wave functions
lead to the same long-distance physics, which is described by the chiral
boson theories for two different hierarchical constructions of the $K$
matrix \cite{kmat}.

The lowest-order Jain hierarchy corresponds to the matrices
$K_{ij} =\pm \delta_{ij} + p\ C_{ij}$, where $p$ is an even, positive integer
and $C_{ij}=1,\ \forall\ i,j=1,\dots,m$ \cite{kmat}. These describe the
experimentally observed filling fractions $\nu=m/(mp\pm 1)$.
For these $K$ matrices, additional currents can be defined
which enlarge the symmetry from $\widehat{U(1)}^{\otimes m}$ to
$\widehat{U(1)}\otimes \widehat{SU(m)}_1$ \cite{read}\cite{kmat},
the last being the non-Abelian current algebra of level one \cite{bpz}.
This specific many-component chiral boson theory explains all the experiments
discussed above, after the inclusion of the interaction with
impurities \cite{kane}.

Nevertheless, in this framework, there is no {\it a-priori} reason why
the most stable hierarchical fluids should have the
$\widehat{SU(m)}_1$ symmetry. Actually, there are also $K$ matrices with
$SO(k)_1$ and $(E_n)_1$ symmetry \cite{froh}, but their filling fractions
do not match well the experimental pattern.
Another weak point is that the concept of composite fermion
has not yet been translated in the language of edge excitations.

\bigskip
\noindent{\bf The $\winf$ symmetry of edge excitations}

In a recent series of papers, we have developed the idea that a specific
symmetry characterizes the Laughlin incompressible fluids
and their edge excitations.
At the classical level, a droplet of incompressible fluid can take different
shapes of the same area, i.e. same density. These configurations are mapped
into each other by reparametrizations of the coordinates of the
plane which preserve the area. This is the {\it dynamical symmetry} of
classical incompressible fluids under {\it area-preserving diffeomorphisms}
\cite{ctz1}\cite{sakita}, whose algebra is called $w_\infty\ $ \cite{shen}.
In particular, the infinitesimal shape changes, which are the classical
edge waves, can be produced by applying the $w_\infty$ infinitesimal
generators to the ground-state droplet configuration \cite{ctz2}.

In our first two papers \cite{ctz1}\cite{ctz2}, we have shown that the quantum
incompressibility of the Laughlin ground states can be expressed as
{\it highest-weight conditions} of the infinite-dimensional $\winf$
algebra, the quantum analogue of $w_\infty$ (see also refs.\cite{flohr}).
Egde excitations are obtained by applying $\winf$ generators with negative
mode index to the ground state.
Moreover, in the thermodynamic limit of infinite particle number
$\left(N\propto R^2\to\infty\right)$, the Laughlin ground state becomes
classical and possesses the $w_\infty$ symmetry according to the above
picture \cite{ctz2}.

In a subsequent paper \cite{cdtz1}, we explicitly constructed the quantum
theory
of edge excitations of the incompressible fluid at integer fillings.
This was achieved by taking the thermodynamic limit of the states near
the edge. The resulting edge theory was shown to correspond to the chiral
boson theory \cite{stone}\cite{wen}.
Moreover, any microscopic interaction of the electrons can be
expanded in the same {\it scaling} limit, leading to a universal form
for the dispersion relation of the edge excitations \cite{ctz4},
which agrees with experiments \cite{wasser}.
As expected, the one-component chiral boson theory is characterized by
the $\winf$ symmetry. In this case,
the $\winf$ algebra is actually generated by polynomials of the edge current;
thus it includes the current algebra $\widehat{U(1)}$ as a subalgebra.

Having identified the correct symmetry for the simplest examples of
incompressible fluids, we proposed to characterize all
quantum incompressible fluids as $(1+1)$-dimensional $\winf$ theories
\cite{ctz3}. The general edge theory can be constructed by
using the algebraic methods of conformal field theory:
the complete Hilbert space of the theory is built by grouping representations
of the symmetry algebra which are closed under the fusion rules, the
composition rules for representations \cite{bpz}.
All unitary, irreducible $\winf$ representations were obtained in the
fundamental work by Kac and collaborators \cite{kac1}\cite{kac2}:
they exist for integer central charge $(c=m)$ and can be regular, i.e.
{\it generic}, or {\it degenerate}.
In ref.\cite{ctz3}, we used the generic representations to build the
{\it generic} $\winf$ theories, which were shown to
correspond to $m$-component chiral boson
theories parametrized by generic $(m\times m)$ $K$ matrices \cite{ctz3}.

\bigskip
\noindent{\bf The content of this paper}

Here, we pursue our study of $\winf$ edge theories by
constructing the special theories made by {\it degenerate}
$\winf$ representations. These theories were not treated in ref. \cite{ctz3}
because the complete mathematical theory of $\winf$ \reps was only made
available afterwards \cite{kac2}.

Degenerate representations are common in conformal
field theory: if the central charge and the weight of a given
representation satisfy certain algebraic relations, some of its
states decouple, and should be projected out
to obtain an irreducible representation.
Degenerate representations form closed sets under the fusion rules,
which are called {\it minimal models}.
There are specific minimal models for any symmetry algebra:
the well-known ones are the $c<1$ Virasoro minimal models \cite{bpz};
larger symmetry algebras, like $\winf$, have $c>1$ minimal models.
The minimal models have less states than the generic theories with
the same symmetry, due to the projection; for the same reason, they
have a richer dynamics.

In this paper, we show that the $\winf$ minimal models correspond to the edge
theories of the Jain hierarchy, which fit the experimental data.
The mathematical rules for building the degenerate representations
have a hierarchical structure similar to the Jain construction:
here, we fully explain this correspondence to the lowest order of the
hierarchies.

This result has far-reaching consequences, both theoretical and experimental.
The physical mechanism which stabilizes the observed quantum Hall fluids
has both short and long distance manifestations.
At the microscopic level, it can be described by the Jain composite-electron
picture and by the size of the gaps;
in the scaling limit, by the minimality of the $\winf$ edge theory.
Actually, we find it rather natural that the theories with a minimal set
of excitations are also dynamically more stable.
This long-distance stability principle leads to a logically self-contained
edge theory of the fractional Hall effect:
a thoroughful derivation of experimental
results is obtained from the principle of $\winf$ symmetry, which is
the basic property of the Laughlin incompressible fluid.

The $\winf$ minimal models are {\it not} realised by the multi-component chiral
boson theories with $\widehat{U(1)}^{\otimes m}$ symmetry, because the latter
do not incorporate the projection for making irreducible the $\winf$
representations of degenerate type.
Nevertheless, reducible and irreducible degenerate representations have
the same quantum numbers of fractional charge, spin and statistics.
The existing experiments at hierarchical filling fractions were
sensible to these data only; therefore their successful interpretation
within the chiral-boson theory is consistent with our theory.
More refined experiments are needed to test the difference.

The $\winf$ minimal theories are realised by the
$\u1\otimes {\cal W}_m(p=\infty)$ conformal theories \cite{kac2}, where the
${\cal W}_m(p)$ are the Zamolodchikov-Fateev-Lykyanov models with
$c=(m-1)\left[1-m(m+1)/p(p+1)\right]$ \cite{fz}.
The main differences with respect to the chiral-boson theory are as follows:

i) There is a {\it single} Abelian current, instead of $m$ independent
ones, and therefore a single elementary (fractionally) charged excitation;
there are neutral excitations, but they cannot
be associated to $(m-1)$ independent edges.

ii) The dynamics of these neutral excitations is new: they have
an $SU(m)$ (not $\widehat{SU(m)}_1$) ``isospin'' quantum number,
because their fusion rules are given by the branching rules of this group.
Therefore, they are quark-like and their fractional statistics
is non-Abelian \cite{moore}.
For example, the edge excitation corresponding to the electron,
for the filling fractions $\nu=m/(mp\pm 1)$, is a
composite made of $(mp)$ anyons and one ``quark'', and carries both
the additive electric charge and the $SU(m)$ isospin.

iii) The degeneracy of particle-hole excitations at fixed angular momentum
is modified by the projection of the minimal models. This counting of states
is provided by the characters of degenerate $\winf$ representations,
which are known \cite{kac2}.
If the neutral $SU(m)$ excitations have a bulk gap, the particle-hole
degeneracy of the ground state (the Wen topological order on the disk
\cite{topord}) is different from the corresponding one of $\u1^{\otimes m}$
excitations.
This can be tested in numerical diagonalizations of few electron
systems; existing data are not accurate enough \cite{wen}.

\medskip

The plan of the paper is the following. In section 2, we
review the generic $\winf$ theories \cite{ctz3}, and their
chiral boson realizations. In section 3, we construct the
$\winf$ minimal models out of degenerate representations.
In section 4, we describe some physical properties of
the minimal models and propose some tests for them.
In the concluding remarks, we discuss the open problem of the higher-order
hierarchical construction of $\winf$ minimal models.

\vbox to .5in {\vfill}
%
%-2------------------------------------------------
%
\section{Existing theories of edge excitations and experiments}

In this section, we shall review the essential points of the {\it generic}
$\winf$ theories of edge excitations \cite{ctz3}.
We shall moreover show in detail how these $\winf$ field theories are
realized as theories of multi-component chiral bosons,
thus making contact with previous treatments \cite{kmat}\cite{wen}
and providing a symmetry ground for them.

\bigskip

\noindent{\bf The generic $\winf$ theories}

At the classical level, a two-dimensional droplet of incompressible
fluid can take different shapes of the same area.
The configuration space of a classical (chiral) incompressible fluid
can thus be generated by {\it area-preserving diffeomorphisms} from
a reference droplet, say a disk \cite{ctz1}\cite{sakita}.
The infinitesimal generators of area-preserving diffeomorphisms satisfy
an infinite-dimensional Lie algebra called $w_\infty$ \cite{shen}.
When they are applied on the classical ground-state distribution function,
they produce small excitations corresponding to (chiral) edge waves
localized on the one-dimensional boundary \cite{ctz2}\footnote{
A complete introduction can be found in section 2 of ref. \cite{ctz4}.}.

This picture is essentially maintained at the quantum level, in the
thermodynamic limit $N\to\infty$.
In this case, the relevant symmetry algebra is $\winf$, the quantum
analogue of $w_\infty$.
The generators $V^i_n$ of $\winf$ are characterized by a mode index
$n \in {\bf Z}$ and an integer {\it conformal spin} $h =i+1 \ge 1$.
They satisfy the algebra \cite{shen},
\beq
{[ V^i_n, V^j_m]} = (jn-im) V^{i+j-1}_{n+m} +q(i,j,n,m)V^{i+j-3}_{n+m}
+\cdots +\delta^{ij}\delta_{n+m,0}\ c\ d(i,n) \ ,
\label{walg}\eeq
where the structure constants $q$ and $d$ are polynomial of their arguments,
$c$ is the central charge, and the dots denote a finite number of
similar terms involving the operators $V^{i+j-2k}_{n+m}\ $
(the complete expression of (\ref{walg}) is a bit cumbersome
and will be given later).
The operators $V^0_n$ satisfy the Abelian current algebra (Kac-Moody algebra)
$\u1$ and the operators $V^1_n$ the Virasoro algebra \cite{bpz}, respectively,
\barr
{[V^0_n,V^0_m]}    & = & n c\ \delta_{n+m,0} \ ,\\
{[ V^1_n, V^0_m }] & = & -m\ V^0_{n+m} \ ,\nonumber\\
{[ V^1_n, V^1_m ]} & = & (n-m)V^1_{n+m} +{c\over 12}n(n^2-1)
\delta_{n+m,0}\ .
\label{walg1}\earr
$V^0_n$ and $V^1_n$ are identified as the charge and angular-momentum modes
of the edge excitations, respectively.
In the classical limit, all terms but the first in the r.h.s. of (\ref{walg})
vanish; the resulting algebra is the classical algebra
$w_\infty$ of area-preserving diffeomorphisms.
Conversely, the algebra (\ref{walg}) is the unique quantization
(up to changes of basis) of $w_\infty$ on the circle \cite{radul}.

A $\winf$ theory is defined as a {\it Hilbert space} constructed as a set of
irreducible, unitary, highest-weight \reps of $\winf$,
which are closed under the fusion rules for making composite states.
This is the well-known algebraic construction of conformal field
theories \cite{bpz}.
No reference to a Hamiltonian governing the dynamics of excitations is needed.
In this formalism, the incompressible quantum fluid ground state appears as
a highest-weight state $|\Omega\rangle_W$ satisfying
\beq
V^i_n \vert\Omega\rangle_W=0\ , \quad \forall\ n >0\ ,\ i\ge 0\ ,
\label{whwc}\eeq
and
\beq
V^i_0 \vert\Omega\rangle_W=0\ ,\ i\ge 0\ .
\label{weige}\eeq
Particle-hole edge excitations above the ground state are obtained by applying
generators with negative mode index to $\vert\Omega\rangle_W$.

The {\it quasi-particle} excitations are the Laughlin anyons \cite{laugh},
which are localized deformations of the bulk density.
Due to incompressibility, their charge excess (or defect)
is entirely transmitted to the boundary, where it is seen as a charged
edge excitation \cite{cdtz1}.
These excitations appear as further highest-weight states,
which have non-vanishing eigenvalues for all the $V^i_0$.
Specifically, $V^0_0$ and $V^1_0$ determine (minus) the charge
and the spin of the quasi-particle, respectively \cite{ctz3}.
The eigenvalues of $V^i_0\ (i\ge 2)$ measure the radial moments of
the charge distribution of quasi-particles \cite{ctz4}.

For integer filling, one can show explicitly that the excitations
above the incompressible ground state can be organized as a $\winf$
representation \cite{ctz1}\cite{cdtz1}.
This fact, combined with the classical picture, has led us
to characterize {\it any} quantum incompressible fluid as a $\winf$ theory.
The hierarchy problem reduces therefore to a complete classification
of $\winf$ theories.

This classification can be achieved thanks to the crucial work \cite{kac1},
in which all irreducible, unitary, {\it quasi-finite} highest-weight
\reps of $\winf$ have been constructed.
Such representations exist only if the central charge is a positive
integer, $c=m$, $m\in {\bf Z}_+$.
They are characterized by a {\it $m$-dimensional weight vector }
$\vec{r}$ with real elements, and are built on top of
a highest weight state $|\vec{r}\rangle_W$, which satisfies,
\beq
V^i_n|\vec{r}\rangle_W = 0 \ ,\qquad \forall\ n>0\ , \ i \ge 0 \ ,
\label{whst}\eeq
and is an eigenstate of the $V^i_0$,
\beq
V^i_0|\vec{r}\rangle_W =\sum_{n=1}^m \ m^i (r_n)\ |\vec{r}\rangle_W
\label{weig}\eeq
where $m^i(r)$ are $i$-th order polynomials of a weight component
(see section 3). In particular, the charge $V^0_0$ and Virasoro $V^1_0$
eigenvalues are given by
\barr
\sum_{n=1}^m \ m^0 (r_n) &=& r_1 + \cdots + r_m\ ,\nonumber\\
\sum_{n=1}^m \ m^1 (r_n) &=& {1\over 2}\left[
  \left(r_1 \right)^2 +\cdots + \left(r_m \right)^2 \right]\ .
\label{wcs}\earr

For {\it generic} $\winf$ \reps, the fusion rules require that all weight
vectors $\vec{r}$ forming a $\winf$ theory span a {\it lattice} $\Gamma$,
\beq
\Gamma=\left\{ \left. \vec{r}\ \right\vert\ \vec{r}=
\sum_{i=1}^m n_i \vec{v}_i\ ,\quad n_i \in {\bf Z} \right\} \ ,
\label{latt}\eeq
whose points satisfy $(r_i-r_j)\not\in {\bf Z},\ \forall\ i\neq j$,
as better explained below.
The resulting $\winf$ theory describes an $m$-component incompressible
fluid with $m$ interacting edges. The $\vec{v}_i$ are identified as the
basis vectors, representing a physical elementary excitation in the
$i$-th component.
The {\it physical charge} of an excitation with labels $n_i\in {\bf Z}$ is
given by the sum of the components in the physical basis
\cite{ctz3}\footnote{
As also explained afterwards in eq. (2.33).},
\beq
Q = \sum_{i,j=1}^m\ K^{-1}_{ij}\ n_j \ ,\qquad n_i \in {\bf Z} \ ,
\label{qform}\eeq
where
$ K^{-1}_{ij} =\left( \vec{v}_i\cdot \vec{v}_j \right) $
is the metric of the lattice. It is related to the highest weight
$\vec{r}$ by a linear basis transformation.
The spin $J$ and the statistics $\theta/\pi$ of this
excitation are given by the Virasoro $V^1_0$ eigenvalue (\ref{wcs}),
\beq
2 J={ \theta\over \pi} = \sum_{i,j=1}^m\ n_i \ K^{-1}_{ij}\ n_j \ ,
\label{thetaform}\eeq
while the Hall conductivity can be computed via the chiral anomaly equation
and is found to be \cite{ctz3}:
\beq
\sigma_H={e^2\over h}\ \nu \ ,\qquad \nu= \sum_{i,j=1}^m\ K^{-1}_{ij} \ .
\label{sigmah}\eeq

The presence of $m$ electron excitations in the spectrum
(\ref{qform},\ref{thetaform}), with unit charge,
fermionic statistics and integer monodromy relative to any other excitation
requires the matrix $K$ to have {\it integer} entries, odd on the diagonal.

Note that there are many bases for the lattice $\Gamma$ (\ref{latt}), which
are related by integer linear transformations, the {\it modular
transformations},
\beq
\vec{u}_i=\sum_{j=1}^m\ \Theta_{ij}\ \vec{v}_j\ , \qquad \Theta_{ij} \in
SL(m,{\bf Z}) \ .
\label{modinv}\eeq
Actually, different bases correspond to different definitions
of the charge units of the fluids, and give different
spectra for the total charge $Q$, while the fractional
statistic spectrum is independent of the basis.

\vbox to .5 in {\vfill}

\noindent{\bf The chiral boson theories}

We now recall the main properties of the
multi-component chiral boson theories \cite{wen}\cite{kmat}.
On an annulus geometry, with edge circles $|{\bf x}|=R_1$ and $|{\bf x}|=R_2$,
one introduces $m$ independent one-dimensional {\it chiral} currents
\beq
J^i \left(R_1\theta-v_i t\right) = - {1\over 2\pi R_1}
{\partial\over\partial\theta} \ \phi^i \ , \qquad (|{\bf x}|=R_1),
\label{jchi}\eeq
and corresponding ones with opposite chirality
$J^i \left(R_2\theta+v_i t\right)$ at the other edge $|{\bf x}|=R_2$.
The dynamics of these currents on the edge circle $|{\rm x}|=R_1$
is governed by the action,
\beq
S=-{1\over 4\pi}\ \int\ dt\ dx\ \sum_{i=1}^m\ \kappa_i
\left(\partial_t\phi^i +v_i \partial_x\phi^i \right)
\partial_x\phi^i\ , \qquad {\rm for}\ \ \  x\equiv R_1\theta\ ,
\label{bosact}\eeq
for the $m$ $(1+1)$-dimensional {\it chiral boson} fields $\phi^i$ \cite{flo}.
The corresponding action for the other circle
$x\equiv R_2\theta$ is obtained by replacing $v_i\to (-v_i)$.
The dynamics on the two edges are identical and independent, only constrained
by the conservation of the total charge: thus we describe one of them only.
We can change the normalization of the fields, and reduce each
coupling constant to a sign, $\kappa_i \to \pm 1$.
The equations of motion imply that the fields are
chiral, $\phi^i=\phi^i(x-v_it)\ $, and canonical quantization
implies the following commutation relations for the currents,
\beq
{[ J^i(x_1),J^k(x_2) ]}= {1\over 2\pi\kappa_i}\ \delta^{ik}\
\delta^{\prime}(x_1-x_2)\ ,\qquad (t_1=t_2) \ ,
\label{curalg}\eeq
which are those of the multi-component Abelian current algebra
$\widehat{U(1)}^{\otimes m}$ \cite{bpz}. The Hamiltonian is
\beq
H=\pi\ \int\ dx\ \sum_{i=1}^m\ \kappa_i v_i\ : J^i J^i :\ ,
\label{ham}\eeq
where the double dots denote normal ordering.
Its positive definiteness requires the signs of the velocities
$v_i$ and the couplings $\kappa_i$ to be related:
\beq
v_i\kappa_i > 0\ , \qquad i=1,\dots ,m\ .
\label{sign}\eeq
Let us first discuss one particular chiral current, $v_i>0$
(i.e. $\kappa_i=1$). The quantization of the chiral boson is equivalent to
the construction of the representations of
the current algebra (\ref{curalg}). Actually, all the states in the
Hilbert space of the theory can be fitted into a set of
representations \cite{bpz}.
To this end, we introduce the Fourier modes of the currents,
\beq
J^i(R\theta-v_it)={1\over 2\pi}\ \sum_{n=-\infty}^{\infty}\
\alpha^i_n\ {\rm e}^{in(\theta-v_it)}\ ,
\label{mod1}\eeq
which satisfy,
\beq
{[}\alpha^i_n,\alpha^j_m {]}= \ \delta^{ij}\ {n\over \kappa_i}\
\delta_{n+m,0}\ .
\label{km}\eeq
The positivity of the ground-state expectation value
\beq
\langle\Omega |\alpha^i_n \alpha^i_{-n} |\Omega\rangle
\equiv \Vert\alpha^i_n |\Omega\rangle\Vert^2\ \ge 0\ ,
\label{schwi}\eeq
and the commutation relations (\ref{km}) with $\kappa_i=1$ imply the conditions
\beq
\alpha^i_n |\Omega\rangle=0\ , \qquad n>0 \qquad\quad (v_i >0).
\label{hwsc}\eeq
An irreducible highest-weight representation of the
$\widehat{U(1)}$ current algebra is made by the highest-weight state
$|\Omega\rangle$ and by all states obtained by applying any
number of $\alpha^i_n, \ \ n<0\ ,$ to it.
The weight of the representation is given by the eigenvalue of $\alpha^i_0$,
which is the single-edge charge, in units to be specified later.
For the ground state, we have
\beq
\alpha^i_0 |\Omega\rangle = 0\ .
\label{cvac}\eeq
Other unitary highest-weight representations can be similarly built on top
of other highest-weight states $|r\rangle$, $r\in {\rm R}$, which satisfy
\beq
\alpha^i_n |r\rangle= 0 \quad n>0\ ,\qquad \alpha^i_0|r\rangle = r
|r\rangle\ ,
\label{hwsr}\eeq
and are built by applying the {\it vertex operators} to the
ground state \cite{bpz}.
These \reps correspond to the quasi-particle excitations of this edge theory.
The Virasoro generators are defined by the Sugawara construction \cite{bpz},
\beq
L^i_n ={\kappa_i\over 2}\ \sum_{l=-\infty}^{\infty}\ :\
\alpha^i_{n-l}\alpha^i_l\ :\ ,
\label{ln}\eeq
and act on the highest-weight states as follows,
\beq
L^i_n |r\rangle =0, \quad n>0\ ,\qquad L^i_0 |r\rangle =
{\kappa_i r^2\over 2} |r\rangle\ .
\label{lnhws}\eeq
They give rise to the Virasoro algebra (\ref{walg1}) with $c=1$, for each
current component $i$.

Let us now discuss an antichiral edge current, $v_i <0$ (i.e. $\kappa_i=-1$).
The modes of the antichiral current are now called $\bar\alpha^i_n$ ,
\beq
J^i(R\theta+v_it)={1\over 2\pi}\ \sum_{n=-\infty}^{\infty}\
\bar\alpha^i_n\ {\rm e}^{in(\theta+v_it)}\ .
\label{mod2}\eeq
The $\bar\alpha^i_n$ satisfy the current algebra (\ref{km}) with
$\kappa_i=-1$. The positivity of (\ref{schwi}) requires in this case
\beq
\bar\alpha^i_n |\Omega\rangle =0\ ,\qquad n<0 \qquad (v_i <0)\ .
\label{antichi}\eeq
The Virasoro generators $\bar{L}^i_n$ are again defined in terms of the
$\bar\alpha^i_n$ by the Sugawara construction (\ref{ln}), and similarly satisfy
$\bar{L}^i_n |\Omega\rangle =0, \ n<0\ $.
These are non-standard ground-state conditions, which also imply that
the operators $\bar{L}_n\ $ do not satisfy the Virasoro algebra (\ref{walg1}).
Nevertheless, it is possible to define operators,
\beq
\alpha^i_n \equiv\bar\alpha^i_{-n}\ ,\qquad
L^i_n \equiv -\bar{L}^i_{-n}\ ,\qquad (v_i<0)\ .
\label{achi}\eeq
which satisfy the standard algebras (\ref{ln},\ref{walg1})
and the standard h.w.s. conditions (\ref{hwsr},\ref{lnhws}) with
$|\kappa_i|=1$.
On the other hand, $\bar{L}^i_0$ retains its physical meaning of
fractional spin and statistics and thus has negative spectrum for $v_i<0$.

The Hamiltonian (\ref{ham}) can be finally expressed in terms of the Virasoro
generators as follows \cite{bpz},
\beq
H= \sum_{i=1}^m \ {|v_i|\over R} \left(L^i_0 -{1\over 24}\right) \ ;
\label{hami}\eeq
its spectrum is positive for both chiralities, as originally required.

Let us now define the unit of charge carried by the $i$-th edge
by coupling the current $J^i$ minimally to an electric field along the edge
$E^i=-\partial A^i_0(\theta,t)/\partial\theta$.
In general, an edge excitation can have components on all edges.
Therefore, we introduce new currents $\rho^i$, carrying one
unit of  charge in the $i$-th edge. These are general linear combinations of
the
$J^i$, which create instead energy eigenstates of definite chirality:
\beq
H_{\rm e.m.}=\int\ dx \ \sum_{i=1}^m \ \rho^i A^i_0 \ , \qquad
\rho^i =\sum_{j=1}^m\ \Lambda_{ij}\ J^j \ ,\qquad \Lambda\in GL(m,{\rm R}) \ .
\label{phyc}\eeq
The total charge is
\beq
Q =  \sum_{i,j=1}^m\ \Lambda_{ij}\ \alpha^j_0 \ ,
\label{qdef}\eeq
In a chiral theory, the minimal coupling (\ref{phyc})
produces the chiral anomaly \cite{bpz}.
This means that the charge is not conserved, i.e. the ground
state evolves into some charged state. By integrating this
evolution over asymptotic times, we can generate all charged
states in the theory, and obtain the spectrum of $\alpha^i_0\ $:
\barr
\left. \Delta \alpha^i_0 \right\vert^{t=\infty}_{t=-\infty} &
= & i\int\ dt\ [\alpha^i_0, H_{\rm e.m.} ] \nonumber\\
& =& {1\over |\kappa_i|2\pi}\sum_{j=1}^m\ \Lambda_{ji} \int\ dt\ dx E^j =
\sum_{j=1}^m\ \Lambda^T_{ij} \ n_j\ ,\qquad n_j \in {\bf Z}\ .
\label{chian}\earr
In this equation, the total integral of the $(1+1)$-dimensional electric
field is a topological invariant quantity, which takes integer values only.
Indeed, this electric field can be produced by adding a solenoid
in the center of the annulus, and by switching on
an integer number of flux quanta \cite{laugh}.
These results show that $\widehat{U(1)}^{\otimes m}$
representations are characterized by weight vectors
$\vec{r}=\{r_1,\dots,r_m\}$, which span the same  lattice $\Gamma$
of the $\winf$ weights (\ref{latt}), with basis
$\left(\vec{v}_i \right)_j =\Lambda_{ij} $.

In conformal field theories, the {\it fusion rules} are the laws
for making composite excitations \cite{bpz}.
For any pair of representations, i.e. of charged excitations, the
representations obtained by fusing them should also be present
in the theory - the representations must form a closed set under the action of
the fusion rules. For the Abelian current algebra, this rule
is simply the addition of weight vectors \cite{bpz}.
Denoting by $M({\bf g}, c, \vec{r})$ the irreducible representations
of the algebra {\bf g} with central charge $c$ and weight $\vec{r}$,
we have
\beq
M \left(\u1^{\otimes m}, m, \vec{r}\right)\bullet
M \left(\u1^{\otimes m}, m, \vec{s}\right) =
M \left(\u1^{\otimes m}, m, \vec{r}+\vec{s}\right) \ .
\label{frule}\eeq
The lattice $\Gamma$ (\ref{latt}) is the closed set for this rule, because
for any pair $\vec{r},\vec{s}\in\Gamma$, the vector $(\vec{r}+\vec{s})$
also belongs to $\Gamma$.

The spectrum of the physical charge and fractional statistics of any edge
excitation is obtained by replacing the spectrum of $\alpha^i_0$
(\ref{chian}) into eqs.(\ref{qdef},\ref{lnhws}):
\barr
Q & = &\sum_{i,j=1}^m\ K^{-1}_{ij}\ n_j \nonumber\\
{ \theta\over \pi} & = &\sum_{i,j=1}^m\ n_i \ K^{-1}_{ij}\ n_j \ ,
\qquad n_i \in {\bf Z} \ ,
\label{qtheta}\earr
where
\beq
K^{-1}_{ij} =\sum_{l=1}^m\ \Lambda_{il} {1\over \kappa_l} \Lambda^T_{lj} =
\left( \vec{v}_i\cdot \eta\cdot \vec{v}_j \right) \ .
\label{met}\eeq
In general, the metric $K^{-1}$ of the lattice $\Gamma$ in the basis
$\vec{v}_i$ is pseudo-Euclidean with signature
$\eta_{ij}=\delta_{ij}\kappa_i$, due to the possible presence of excitations
with different chiralities.

The Hall conductivity in the annulus geometry can be measured by applying
a uniform electric field along all the edges, $E^i=E$.
The chiral anomaly of the edge theory actually corresponds to a radial
flow of particles in the annulus, which move from the inner edge
to the outer edge. From eq.(\ref{chian}), the Hall conductivity can be thus
identified as
\beq
\sigma_H={e^2\over h}\ \nu \ ,\qquad \nu= \sum_{i,j=1}^m\ K^{-1}_{ij} \ .
\label{sig}\eeq
Equations (\ref{qtheta},\ref{met},\ref{sig}) for the Hall conductivity and
the spectrum of the charge and fractional statistics of the chiral boson
theories reproduce those of the generic $\winf$ theories
(\ref{qform},\ref{thetaform},\ref{sigmah}).
As before, the existence of $m$ electron excitations with unit charge
and integer statistic relative to all excitations,  requires that
$K$ has integer entries with odd integers on the diagonal.

Note that the action (\ref{bosact}) can be also presented in the basis of
charge eigenstates $\rho_i$,
\beq
S=-{1\over 4\pi}\ \int\ dt\ dx\ \sum_{i,j=1}^m\ \left(
K_{ij} \partial_t\tilde\phi^i +V_{ij} \partial_x\tilde\phi^i \right)
\partial_x\tilde\phi^j\ ,
\label{newact}\eeq
where we define $\rho^i = -\partial_x \tilde{\phi}^i /2\pi\ $,
and $V_{ij} =\sum_{l=1}^m\ \Lambda^T_{il}\ |v_l|\ \Lambda_{lj}$.
Note that $K$ can be any symmetric matrix, while $V$ has a specific form.
More general forms of $V$ would give a non-diagonal
Hamiltonian in the previous basis.
These interactions among the edges have been discussed in ref.\cite{kane}.
Let us remark that the quantities $Q, \theta/\pi $ and $ \nu$ derived
before are  {\it universal kinematical data} of the quantum incompressible
fluid, which are independent of the dynamical data encoded in $V$, i.e. in
the Hamiltonian. Actually, only $H_{\rm e.m.}$ (\ref{phyc}) and the
current algebra (\ref{km}) were used in their derivation.
Other physical quantities can depend on $V$, like the time dependence of the
correlators and the Hall conductivity in the bar geometry \cite{wen}.
On the other hand, these edge interactions were shown to be irrelevant,
due to the random effect of disorder, when $K$ describes the physical
Jain fluids \cite{kane}, which we shall deal with in the following.

\bigskip

\noindent{\bf $\winf$ theories and chiral boson theories}

We have introduced two kinds of edge theories, the generic $\winf$
theories and the chiral boson theories, which have the same spectra of filling
fractions and fractional charge and statistics of the excitations.
We now show that:

i) the generic $\winf$ theories are equivalent to chiral boson theories;

ii) there are {\it more} $\winf$ symmetric theories, which are different.

In the algebraic approach, the first point is proven by identifying
the generic $\winf$ \reps with $\u1^{\otimes m}$ \reps
- we must qualify the word ``generic''.
We already saw that both \reps are labelled by the same weight vectors
$\vec{r}$. We must also map one-to-one the states built on top of the
respective highest weight states.

The general theory of unitary, irreducible (quasi-finite) $\winf$ \reps,
developed in refs. \cite{kac1}\cite{kac2}, leads to the following relations
between irreducible \reps of the two algebras,
\beq\begin{array}{l l l}
M\left(\winf ,1,r\right) &\sim& M\left(\widehat{U(1)},1,r\right)\ ;\\
M\left(\winf ,m>1,\vec{r}\right) &\sim&
M\left(\widehat{U(1)}^{\otimes m},m,\vec{r}\right)\ ,
\qquad {\rm for}\ (r_i-r_j) \not\in {\bf Z} \ ,\forall\ i\neq j\ , \\
M\left(\winf ,m>1,\vec{r}\right) & \subset &
M\left(\widehat{U(1)}^{\otimes m},m,\vec{r} \right)\ ,
\qquad {\rm if \ }\exists\ (r_i - r_j) \in {\bf Z} \ .
\end{array}\label{winclu}\eeq
Generically, $\winf$ and $\u1^{\otimes m}$ \reps are one-to-one equivalent.
The exceptions appear for $c>1$, when the weight has some integer
components $(r_i - r_j)$. In these cases, the relation is many-to-one, i.e. an
irreducible $\widehat{U(1)}^{\otimes m}$ representation is {\it reducible}
with respect to the $\winf$ algebra.
We call {\it generic} the $\winf$ \reps which are one-to-one equivalent to
$\u1^{\otimes m}$ ones ($(r_i-r_j) \not\in {\bf Z} \ ,\forall\ i\neq j$),
and {\it degenerate} the remaining $\winf$ \reps
($\exists\ (r_i - r_j) \in {\bf Z}$).

The results (\ref{winclu}) allow the construction of several
types of $\winf$ symmetric theories.
The generic $\winf$ theories \cite{ctz3} are defined by lattices $\Gamma$
(\ref{latt}) which contain generic $\winf$ \reps only:
for these, the basis vectors satisfy
\hbox{ $((\vec{v}_\alpha)_i -(\vec{v}_\alpha)_j)\not\in {\bf Q}$},
$\forall\ \alpha, i\neq j =1,\dots, m$.
These theories are thus equivalent to chiral boson theories\footnote{
The ground state representation ($\vec{r}=0$) must also be a $\u1^{\otimes m}$
\rep for the closure of the fusion rules.}.
Other $\winf$ theories, containing only degenerate \reps, are instead
different. These are the {\it minimal models}, which we shall explicitly
construct in the next section. They are the basic new $\winf$ theories,
and are actually very important, because they will be shown to
correspond to the experimentally observed Jain fluids.

On the other hand, the chiral boson theories of the Jain hierarchy
have been widely used in the literature and partially confirmed by
the experiments, as we discuss hereafter.
These are also $\winf$ symmetric, but are not the simplest realizations of
this symmetry, because their $\u1^{\otimes m}$ \reps are reducible.
The $\winf$ minimal models will be shown to have basically the same spectrum
of excitations, but to differ in very important properties.

\bigskip

\noindent{\bf The Jain hierarchy}

The Jain fluids have been described by the subset of the chiral boson theories
characterized by the following $K$ matrices \cite{kmat},
\beq
K_{ij}=\pm\delta_{ij} + p\ C_{ij}\ , \qquad C_{ij}=1\ \forall\ i,j=1,\dots,m\ ,
\ p>0\ {\rm even}\ ,
\label{kjain}\eeq
and the following spectra of edge excitations (eqs.(\ref{qtheta},\ref{sig})),
\barr
\nu &= &{m\over mp \pm 1}\ , \qquad p >0 \ {\rm even}\ ,\qquad (c =m)\ ,
\nonumber\\
Q & = & {1\over pm \pm 1}\ \sum_{i=1}^m n_i \ ,\nonumber\\
{ \theta\over\pi} &=& \pm\left( \sum_{i=1}^m n^2_i -
{p\over mp \pm 1}\left( \sum_{i=1}^m n_i \right)^2 \right) \ .
\label{jspec}\earr
Note that $K$ has $(m-1)$ degenerate eigenvalues $\lambda_i =1$,
$\ i=1,\dots, m-1\ $ (resp. $\lambda_i=-1$),
and a single value $\lambda_m=\pm 1+mp$. If the sign $\pm$ is negative,
one edge has opposite chirality to the others.
There is one basic charged quasi-particle excitation with label
$n_i=(1,\dots,1)$ and $m(m-1)/2$ {\it neutral} excitations for
$n_i=(\delta_{ik}-\delta_{il})$, $\ 1\le k <l\le m$, with identical integer
statistics.

The corresponding trial wave functions for the ground state
have been constructed by Jain as \cite{jain}
\beq
\Psi_{\nu}=D^{p/2}\ L^m\ {\bf 1}\ , \qquad p\ {\rm even} \ ,
\label{jjain}\eeq
where $L^m\ {\bf 1}$ represents schematically the wave function of $m$ filled
Landau levels and $D^{p/2}$ multiplies the wave
function by the $p$-th power of the Vandermonde determinant, which
``attaches $p$ flux tubes to each electron'', and transforms them
into ``composite fermions''.
This construction has been implemented in the multi-component
chiral boson theory (\ref{newact}) in refs. \cite{jtrans}\cite{kmat}.

The Jain hierarchy covers most of the experimentally observed plateaus,
as we discuss in the next section.
However, within the chiral boson approach, there is no clear motivation for
selecting the special $K$ matrices (\ref{kjain}).
The size of the gap for bulk density waves is usually invoked for
solving this puzzle: the observed fluids are supposed
to have the largest gaps, while the general $K$ fluids have small gaps
and are destroyed by thermal fluctuations and other effects.
It is also true that edge theories give kinematically possible incompressible
fluids and their universal properties, but cannot describe
the size of the gaps, which is determined by the microscopic
bulk dynamics.

Nevertheless, in this paper, we show that there is a natural way to select
the Jain hierarchy within the $\winf$ edge theory approach.
Indeed, the Jain fluids correspond to the $\winf$ minimal models, which are
characterized by possessing less states than their chiral boson counterparts.
We propose this reduction of available states as a natural stability
principle.

\bigskip

\bigskip

\bigskip

\noindent{\bf Experiments}

We first discuss the spectrum of fractional Hall conductivities in
eq. (\ref{jspec}). According to Jain, the stability of the ground states
(\ref{jjain}) should be approximately independent of $m$, which counts the
number of Landau levels filled by the composite fermion.
This is in analogy with the integer Hall plateaus, which are all
equally stable. On the other hand, the stability decreases
by increasing $|p|$, as observed for the Laughlin fluids ($m=1$).
Therefore, the most stable family of plateaus is,
\beq
\nu={m\over 2m \pm 1} \ ,\qquad (p=2)\ ,
\label{fam1}\eeq
which accumulate at $\nu=1/2$. The next less stable family is
\beq
\nu={m\over 4m \pm 1} \ ,\qquad (p=4)\ ,
\label{fam2}\eeq
which accumulate at $\nu=1/4$.
This behaviour is clearly seen in the experimental data of fig. 1.
For these filling fractions, the Jain wave functions for the ground state
and the simplest excited states have a good overlap with those obtained
numerically by diagonalizing the microscopic Hamiltonian
with a small number of electrons.
Another confirmation of the composite fermion picture comes from the
independent dynamics of the {\it compressible} fluid at $\nu=1/2$:
a theory of weakly interacting composite fermions has been proposed \cite{hlr},
which has been confirmed by the experiments \cite{nu1/2}.

A closer look into the experimental values of the filling fractions
in fig. 1 shows other points (in italic), like $\nu=4/5,\ 5/7,\ 8/11$,
which fall outside the  Jain main series (\ref{fam1},\ref{fam2}) (in bold).
These points were originally interpreted as ``charge conjugates'' of these
series \cite{jain},
\beq
\nu=1-{m\over 2m \pm 1} \ , \qquad \nu=1 -{m\over 4m \pm 1} \ .
\label{ccon}\eeq
A charge-conjugated fluid is a fluid of holes in a $(\nu=1)$ electron fluid.
The corresponding $K$ matrix is easily obtained as the
$((m+1)\times (m+1))$-dimensional matrix \cite{kmat},
\beq
\bar{K}=  \left( \begin{array}{c c} 1 & 0 \\ 0 & -K \end{array} \right)
\label{conj}\eeq
These charge-conjugate models actually belong to the second
iteration of the Jain hierarchy \cite{jain}.
Unfortunately, the charge conjugate states do not fit well the data in
fig. 1. The $\nu=1/2$ family would be self-conjugate; thus there should
be two fluids per filling fraction, which are not observed, apart from two
cases.
Actually, coexisting fluids can be detected by experiments where the magnetic
field is tilted from the orthogonal direction to the plane \cite{tilt}.
Furthermore, the conjugate of the observed fractions in the
$\nu=1/4$ family are not observed in half of the cases.
Finally, there are fractions which do not belong to any previous group:
$\nu=4/11,\ 7/11,\ 4/13,\ 8/13$, $\ 9/13,\ 10/17$.

In conclusion, all the fractions outside the main Jain families
(\ref{fam1},\ref{fam2}) are not well understood at present (and will not be
explained in this paper).
Any known extension of the previous theory which explains these few extra
fractions, also introduces many more unobserved fractions, with an unclear
pattern of stability.
Besides the second iteration of the Jain hierarchy \cite{jain}, we also quote
the approach proposed by Fr\"ohlich and collaborators \cite{froh}.
They analyzed all lattices $\Gamma$ (\ref{latt}), with positive-definite,
integer (inverse) metric $K$, for small values of $det(K)$, whose
classification is known in the mathematical literature. These lattices can
be related to the $SU(m)$, $SO(k)$ and exceptional Lie algebras.
The stability of the corresponding fluids does not follow a clear pattern
related to these algebras, besides the case of the chiral Jain fluids
(\ref{kjain}), whose $SU(m)$ symmetry will be explained in the next section.
Moreover, in this approach, the $K$ matrices for the Jain filling fractions
$\nu=m/(mp-1)\ ,p>0$, are different from the Jain proposal (\ref{kjain})
which is not positive definite.

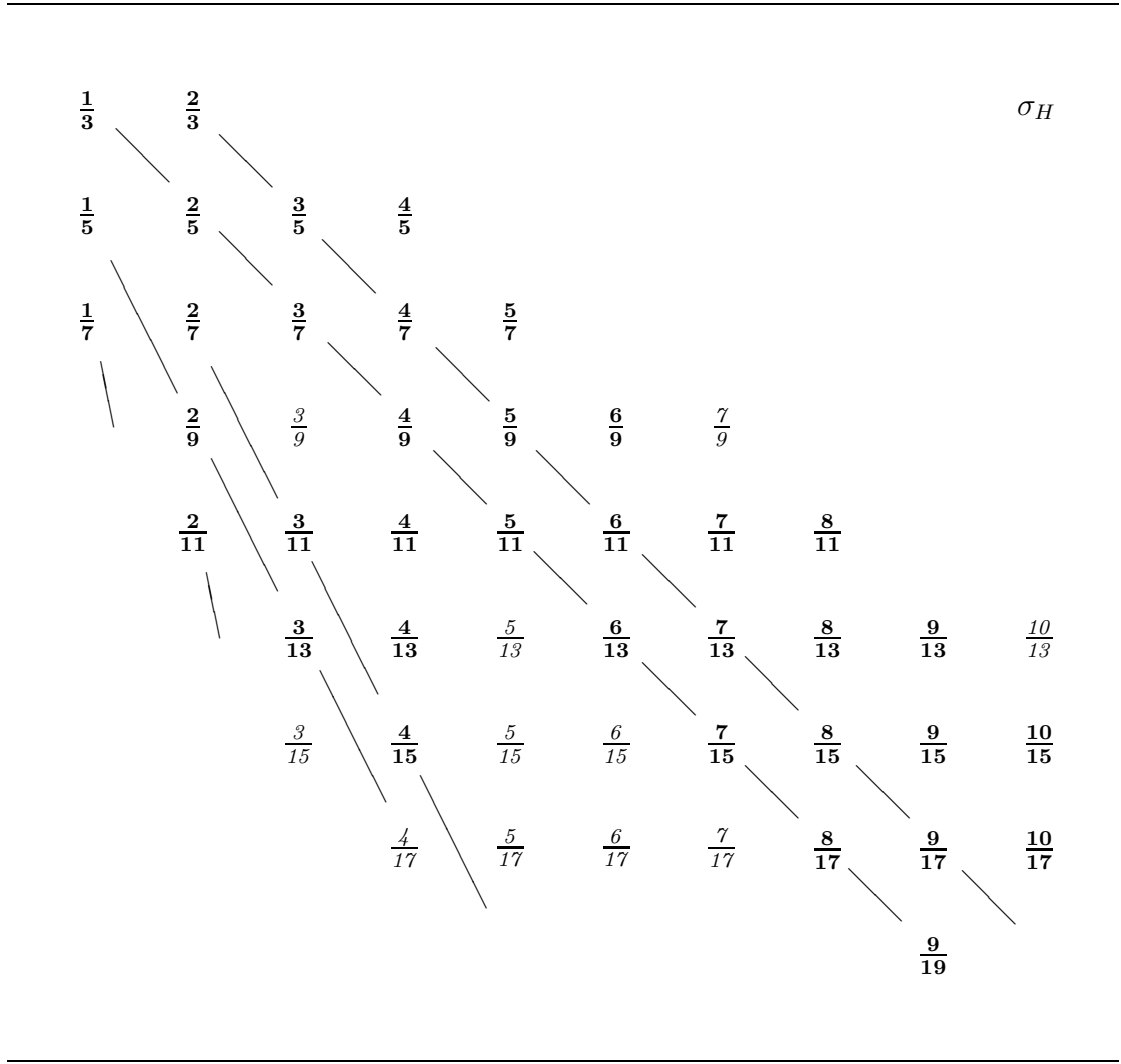
\begin{figure}
\unitlength=1.0pt
\begin{picture}(440.00,440.00)(0.00,0.00)
\put(10.00,500.00){\line(1,0){420.00}}
\put(10.00,100.00){\line(1,0){420.00}}
\put(348.00,153.00){\line(-1,1){20.00}}
\put(309.00,192.00){\line(-1,1){20.00}}
\put(270.00,231.00){\line(-1,1){20.00}}
\put(229.00,273.00){\line(-1,1){20.00}}
\put(191.00,311.00){\line(-1,1){20.00}}
\put(151.00,352.00){\line(-1,1){20.00}}
\put(110.00,394.00){\line(-1,1){20.00}}
\put(71.00,433.00){\line(-1,1){20.00}}
\put(351.00,192.00){\line(-1,1){20.00}}
\put(309.00,233.00){\line(-1,1){20.00}}
\put(270.00,272.00){\line(-1,1){20.00}}
\put(230.00,311.00){\line(-1,1){20.00}}
\put(192.00,350.00){\line(-1,1){20.00}}
\put(149.00,391.00){\line(-1,1){20.00}}
\put(110.00,431.00){\line(-1,1){20.00}}
\put(391.00,152.00){\line(-1,1){20.00}}
\put(153.00,198.00){\line(-1,2){25.00}}
\put(112.00,278.00){\line(-1,2){25.00}}
\put(74.00,353.00){\line(-1,2){25.00}}
\put(112.00,313.00){\line(-1,2){25.00}}
\put(150.00,239.00){\line(-1,2){25.00}}
\put(191.00,158.00){\line(-1,2){25.00}}
\put(50.00,340.00){\line(-1,5){5.00}}
\put(90.00,260.00){\line(-1,5){5.00}}
\put(40.00,460.00){\makebox(0,0)[cc]{$ {\bf 1\over 3}$}}
\put(80.00,460.00){\makebox(0,0)[cc]{$ {\bf 2\over 3}$}}
\put(40.00,420.00){\makebox(0,0)[cc]{$ {\bf 1\over 5}$}}
\put(80.00,420.00){\makebox(0,0)[cc]{$ {\bf 2\over 5}$}}
\put(120.00,420.00){\makebox(0,0)[cc]{$ {\bf 3\over 5}$}}
\put(160.00,420.00){\makebox(0,0)[cc]{$ {\bf 4\over 5}$}}
\put(40.00,380.00){\makebox(0,0)[cc]{$  {\bf 1\over 7}$}}
\put(80.00,380.00){\makebox(0,0)[cc]{$ {\bf 2\over 7}$}}
\put(120.00,380.00){\makebox(0,0)[cc]{$ {\bf 3\over 7}$}}
\put(160.00,380.00){\makebox(0,0)[cc]{$ {\bf 4\over 7}$}}
\put(200.00,380.00){\makebox(0,0)[cc]{$ {\bf 5\over 7}$}}
\put(80.00,340.00){\makebox(0,0)[cc]{$ {\bf 2\over 9}$}}
\put(120.00,340.00){\makebox(0,0)[cc]{$ {\it  3\over 9}$}}
\put(160.00,340.00){\makebox(0,0)[cc]{$ {\bf 4\over 9}$}}
\put(200.00,340.00){\makebox(0,0)[cc]{$ {\bf 5\over 9}$}}
\put(240.00,340.00){\makebox(0,0)[cc]{$ {\bf 6\over 9}$}}
\put(280.00,340.00){\makebox(0,0)[cc]{$ {\it  7\over 9}$}}
\put(80.00,300.00){\makebox(0,0)[cc]{$  {\bf 2\over 11}$}}
\put(120.00,300.00){\makebox(0,0)[cc]{$ {\bf 3\over 11}$}}
\put(160.00,300.00){\makebox(0,0)[cc]{$ {\bf 4\over 11}$}}
\put(200.00,300.00){\makebox(0,0)[cc]{$ {\bf 5\over 11}$}}
\put(240.00,300.00){\makebox(0,0)[cc]{$ {\bf 6\over 11}$}}
\put(280.00,300.00){\makebox(0,0)[cc]{$ {\bf 7\over 11}$}}
\put(320.00,300.00){\makebox(0,0)[cc]{$ {\bf 8\over 11}$}}
\put(120.00,260.00){\makebox(0,0)[cc]{$ {\bf 3\over 13}$}}
\put(160.00,260.00){\makebox(0,0)[cc]{$ {\bf 4\over 13}$}}
\put(200.00,260.00){\makebox(0,0)[cc]{$ {\it  5\over 13}$}}
\put(240.00,260.00){\makebox(0,0)[cc]{$ {\bf 6\over 13}$}}
\put(280.00,260.00){\makebox(0,0)[cc]{$ {\bf 7\over 13}$}}
\put(320.00,260.00){\makebox(0,0)[cc]{$ {\bf 8\over 13}$}}
\put(360.00,260.00){\makebox(0,0)[cc]{$ {\bf 9\over 13}$}}
\put(400.00,260.00){\makebox(0,0)[cc]{$ {\it  10\over 13}$}}
\put(120.00,220.00){\makebox(0,0)[cc]{$ {\it  3\over 15}$}}
\put(160.00,220.00){\makebox(0,0)[cc]{$ {\bf 4\over 15}$}}
\put(200.00,220.00){\makebox(0,0)[cc]{$ {\it  5\over 15}$}}
\put(240.00,220.00){\makebox(0,0)[cc]{$ {\it  6\over 15}$}}
\put(280.00,220.00){\makebox(0,0)[cc]{$ {\bf 7\over 15}$}}
\put(320.00,220.00){\makebox(0,0)[cc]{$ {\bf 8\over 15}$}}
\put(360.00,220.00){\makebox(0,0)[cc]{$ {\bf 9\over 15}$}}
\put(400.00,220.00){\makebox(0,0)[cc]{$ {\bf 10\over 15}$}}
\put(160.00,180.00){\makebox(0,0)[cc]{$ {\it  4\over 17}$}}
\put(200.00,180.00){\makebox(0,0)[cc]{$ {\it  5\over 17}$}}
\put(240.00,180.00){\makebox(0,0)[cc]{$ {\it  6\over 17}$}}
\put(280.00,180.00){\makebox(0,0)[cc]{$ {\it  7\over 17}$}}
\put(320.00,180.00){\makebox(0,0)[cc]{$ {\bf 8\over 17}$}}
\put(360.00,180.00){\makebox(0,0)[cc]{$ {\bf 9\over 17}$}}
\put(400.00,180.00){\makebox(0,0)[cc]{$ {\bf 10\over 17}$}}
\put(360.00,140.00){\makebox(0,0)[cc]{$ {\bf 9\over 19}$}}
\put(430.00,460.00){\makebox(0,0)[rc]{$\sigma_H \qquad $}}
\end{picture}
\caption{
List of all fractions $\nu=p/q$, with $2/11<\nu<4/5$, $1<p\le 10$ and
$3\le q\le 17$, $\ q$ odd.
The fractions corresponding to experimental values of the Hall
conductivity $\sigma_H=(e^2/h)\nu$ are in {\bf bold}; the unobserved fractions
are in {\it italic}. Observed fractions joined by lines
are explained by the Jain hierarchy (2.43,2.44).}
\end{figure}

In figure 2, we study the limitations of phenomenological descriptions
of the stability of the fluids. Besides all the observed (bold)
fractions of fig. 1, we report the unobserved (italic) ones $\nu=p/q$,
which satisfy the conservative cuts of the ``phase space''
\beq
{2\over 11} < \nu={p\over q} < {4\over 5} \ , \qquad {\rm and}\ \
p \le 10\ , q\le 17 \ .
\label{phasp}\eeq
Namely, we display all fractions which would be observed if
the gap were a smooth function of the parameters $(\nu,p,q)$ interpolating
the data, a typical phenomenological hypothesis.
Figure 2 shows that, besides the families (\ref{fam1},\ref{fam2}), about
half of the fractions are unexplained observed fillings
and half are unobserved but phenomenologically possible.
This implies that the gap is not a smooth function of simple parameters
like $(\nu,p,q)$ - deeper theories are needed to explain stability.

Actually, a major virtue of the Jain hierarchy is that of representing
one-parameter families of Hall states, within which the gap {\it is}
a smooth function of the above parameters.
We think of these families as the set of {\it kinematically allowed}
quantum incompressible fluids (at first level of the hierarchy).

More specific confirmations of the edge theory (\ref{kjain}) come
from the experimental tests of the spectrum of excitations (\ref{jspec}).
An experiment with high time resolution \cite{tdom} has measured the
propagation
of a single chiral charge excitation for $\nu=1/3\ $ ($m=1\ ,p=2$),
and $\nu=2/3\ $ ($m=2,\ p=2$); this is in agreement with the Jain theory,
although the neutral excitations have not been seen yet.
The resonant tunnelling experiment \cite{milli} has verified the conformal
dimensions (\ref{jspec}) for the simplest Laughlin fluid $\nu=1/3$ \cite{tunn}.
Extensions of this experiment to $(m>1)$ fluids have been suggested, as well
as tests of the neutral edge spectrum \cite{kane}.
We shall discuss more these experiments in section 4.

Let us finally remark that experiments support the
$\winf$ symmetry of the edge excitations,  because
there is no evidence for more general conformal field theories without
this symmetry, like orbifolds and coset models \cite{bpz}.
Actually, these could easily be obtained by adding degrees
of freedom to the edge excitations.
Such generalizations are not relevant to the fractional Hall effect
for spin-polarized, single-layer electrons.
%%

%\vfill \eject
%\vbox to 1.in {\vfill}

%-3-------------------------------

\section{$\winf$ minimal models}

In this section, we review the work of ref. \cite{kac2} on degenerate
$\winf$ representations and use these to build the $\winf$ minimal models,
which are then shown to correspond to the Jain fluids \cite{jain}.

\bigskip

\noindent{\bf Degenerate representations}

The complete $\winf$ algebra can be given in compact
form by using a parametric sum of the $V^i_n$ current modes, denoted by
$V\left(- z^n \exp (\lambda D)\right)$, where
$D\equiv z{\partial\over \partial z}$ \cite{kac1}.
This satisfies the algebra,
\barr
{[V\left(-z^r {\rm e}^{\lambda D} \right)\ ,\
V\left(-z^s {\rm e}^{\mu D} \right) ]} &= &
\left( {\rm e}^{\mu r} -{\rm e}^{\lambda s} \right)
\ V\left(- z^{r+s}\ {\rm e}^{(\lambda +\mu) D} \right)\ \nonumber\\
&\ & + \delta_{r+s,0}\ c \ {{\rm e}^{-\lambda r} -{\rm e}^{-\mu s} \over
 1 - {\rm e}^{\lambda +\mu} } \ .
\label{wconp}\earr
The currents $V^i_n$ are identified by an expansion of this
parametric operator in $\lambda$,
\beq
V^i_n \equiv V \left(- z^n f^i_n (D) \right) \ ,
\label{wexp}\eeq
where $f^i_n (D)$ are specific $i$-th order polynomials which
diagonalize the central term of (\ref{wconp}) in the $i,j$ indices
\cite{kac1}\cite{ctz4}. For example,
\beq
V^0_n\equiv V\left(-z^n\right)\ ,\qquad
V^1_n \equiv V\left( -z^n \left(D+{n+1\over 2}\right)\right)\ .
\label{vex}\eeq
The unitary irreducible {\it quasi-finite} highest-weight representations
\cite{kac2}, denoted by $M\left(\winf,c,\vec{r}\right)$, exist for $c=m$ and
are characterized by the highest weight state $|\vec{r}\rangle_W$,
which satisfies
\beq
V\left(- z^n {\rm e}^{\lambda D}\right) \vert \vec{r}\rangle_W =0\ ,
\qquad n>0\ ,
\label{whwcc}\eeq
and
\beq
V\left( - {\rm e}^{\lambda D}\right) \vert\vec{r}\rangle_W =
\Delta(\lambda)\vert\vec{r}\rangle_W \equiv
\sum_{i=1}^m \ {{\rm e}^{\lambda r_i} -1 \over {\rm e}^\lambda -1 } \
\vert\vec{r}\rangle_W\ ,
\label{weigen}\eeq
where $\vec{r}=\{r_1,\dots,r_m\}\in {\bf R}$.
In particular, the charge $V^0_0$ and Virasoro $V^1_0$ eigenvalues given
before in (\ref{wcs}) are recovered by expanding $\Delta(\lambda)$
and comparing to (\ref{vex}).

The infinite tower of states in each representation is generated by
expansion in the $\{\lambda_i\}$ of
\beq
V\left(- z^{-n_1} {\rm e}^{\lambda_1 D} \right) \cdots
V\left(- z^{-n_k} {\rm e}^{\lambda_k D} \right) \vert\vec{r}\rangle_W\ ,
\qquad n_1 \ge n_2 \cdots \ge n_k >0 \ ,
\label{wverma}\eeq
where $n=\sum_{i=1}^k n_i $ is the {\it level} of the states.
The quasi-finite representations have only a finite number of independent
states at each level, thus there are an infinity of polynomial relations
among the generators $V^k_n$, whose explicit form depends on the
values of $c$ and $\vec{r}$.
The number of independent states $d(n)$ at level $n$ is encoded in the
(specialized) character of the representation ($\vert q\vert<1 $) \cite{bpz},
\beq
\chi_{M(\winf,m,\vec{r})} (q) \equiv {\rm tr}_{M(\winf,m,\vec{r})}
\left( q^{\scr{V^1_0-{m\over 24}}} \right) =
q^{\scr{\sum_{i=1}^m \left({r_i^2\over 2} -{1\over 24}\right)} }
 \sum_{n=0}^\infty d(n) q^n\ .
\label{wchar}\eeq
We call the representation {\it generic} if the weight $\vec{r}$
has components $(r_i -r_j ) \not\in {\bf Z}\ , \ \forall i\neq j$,
and {\it degenerate} if it has $(r_i-r_j) \in {\bf Z}$ for some $i\neq j$.
The weight components $\{r_i\}$ of the degenerate representations
can be grouped and ordered in {\it congruence classes} modulo ${\bf Z}$
\cite{kac2},
\barr
&& \{ r_i,\dots,r_m \}=\{ s_1+n^{(1)}_1,\dots,s_1+n^{(1)}_{m_1} \} \cup \cdots
\cup \{ s_k+n^{(k)}_1,\dots,s_k+n^{(k)}_{m_k} \}\ ,\nonumber\\
&& n^{(i)}_j \in {\bf Z}\ ,\ n^{(i)}_1 \ge n^{(i)}_2 \ge\cdots
\ge n^{(i)}_{m_i} \ ,\ m=\sum_{i=1}^k m_i \ ,\ s_i \in {\rm R}\ .
\label{congu}\earr
A representation with two classes is the tensor product of two one-class
representations. Therefore, the one-class degenerate
representations are the basic building blocks, which we shall use for
the $\winf$ minimal models.

The character for the generic representations is
\beq
\chi_{M(\winf,m,\vec{r})} (q) = \prod_{i=1}^m \ {q^{r^i/2}\over\eta(q)} =
\prod_{i=1}^m \ \chi_{M(\widehat{U(1)},1,r_i)} (q) \ ,
\label{chir}\eeq
where $\eta(q)$ is the Dedekind function \cite{bpz},
\beq
\eta(q)=q^{1/24} \prod_{n=1}^\infty \left(1-q^n\right)\ .
\label{dede}\eeq
Equation (\ref{chir}) also shows the expression of the $\winf$
character in terms of $m$ characters of the $\widehat{U(1)}$ algebra,
which confirms the one-to-one equivalence of generic $\winf$ and
$\widehat{U(1)}^{\otimes m}$ representations in eq.(\ref{winclu}).

The character for the one-class degenerate representations is \cite{kac2}
\barr
\chi_{M(\winf,m,\vec{r})} (q) &=& \eta(q)^{-m}\ q^{\sum_{i=1}^m r_i^2/2}
\prod_{1\le i<j\le m} \left( 1-q^{n_i-n_j+j-i} \right)\ ,\nonumber\\
\vec{r} = \{r_1,\dots,r_m\} &=& \{s+n_1,\dots,s+n_m\}\ ,\quad
n_1\ge\cdots \ge n_m\ .
\label{chid}\earr
Note that the number of independent states $d(n)$ at level $n$ is lower for
degenerate representations (\ref{chid}) than for generic ones (\ref{wchar}),
because the former have additional relations among the states,
leading to {\it null vectors} \cite{bpz}.
This is the origin of reducibility of the $\widehat{U(1)}^{\otimes m}$
representations with respect to the $\winf$ algebra.
On the other hand, the one-class degenerate $\winf$ representations are
one-to-one equivalent to those of the
$\widehat{U(1)}\otimes {\cal W}_m (p=\infty)$ minimal models, where
${\cal W}_m$ is the Fateev-Lykyanov-Zamolodchikov algebra \cite{fz}.
The ${\cal W}_m(p)$ minimal models exist for the values of the  central charge
\beq
c=(m-1)\left( 1-{m(m+1)\over p(p+1)} \right)\ , \qquad p> m \ge 2\ ,
\label{cwm}\eeq
therefore we are interested in their limit $p=\infty$.

\bigskip

\noindent{\bf The $c=2$ case}

The nature of the $\winf$ degenerate representations and their
equivalence to the $\widehat{U(1)}\otimes {\cal W}_m$ ones\footnote{
{}From now on, we simply denote by ${\cal W}_m$ the $p\to\infty$ limit
of ${\cal W}_m(p)$.}
can be understood in simple terms for $m=2$, where the ${\cal W}_2$ algebra
is the $c=1$ Virasoro algebra.
By explicit construction, we shall derive the relations
\barr
M \left(\winf,2,\{r+n,r\} \right) & \sim &
M \left(\widehat{U(1)}\otimes {\rm Vir},2,
     \{ {2r+n\over\sqrt{2}} \}, \{ {n^2\over 4} \} \right) \nonumber\\
   & {} &\subset M \left(\widehat{U(1)}^{\otimes 2},2,
     \{ {2r+n\over\sqrt{2}} , {n\over\sqrt{2}} \} \right)  \ ,
\label{rep2}\earr
where we characterized the Virasoro \reps by the $L_0$ eigenvalue.

For $c=1$, the $\winf$ algebra is the enveloping algebra of the $\u1$
algebra, i.e. all $V^i_n$ can be written as polynomials of one current mode
$\alpha_n$ (\ref{km}) \cite{cdtz1}\cite{ctz4}.
Thus, the states of a $\winf$ \rep (\ref{wverma}) can be built
by applying any number of $\alpha_n$, with $n<0$. Their degeneracy $d(n)$
is thus equal to the number of partitions of $n$, whose generating
function is the $U(1)$ character in (\ref{chir}).
Given the linearity in $c$ of the $\winf$ algebra,
a pair of current modes $\alpha^1_n$ and $\alpha^2_n$ similarly build
$\winf$ representations for $c=2$.
The $V^0_n$ and $V^1_n$ generators can be written,
\beq
V^0_n=\alpha^1_n+\alpha^2_n\ ,\quad
V^1_n= {1\over 2} \sum_{l=-\infty}^\infty \ : \alpha^1_{n-l}\alpha^1_l :\ +
{1\over 2} \sum_{l=-\infty}^\infty \ : \alpha^2_{n-l}\alpha^2_l :\ .
\label{reg2}\eeq
The degeneracy of states in these $\winf$ \reps is
given by the product of two $\u1$ characters.
Therefore, this construction is useful to show the equivalence of
generic $\winf$ representations with $\widehat{U(1)}^{\otimes 2}$ ones,
by eq. (\ref{chir}).

The degenerate $\winf$ representations are obtained by another construction,
\beq
V^0_n=\sqrt{2}\ \alpha_n\ ,\quad
V^1_n= {1\over 2} \sum_{l=-\infty}^\infty \ : \alpha_{n-l}\alpha_l :\ +
\ L_n \ ,
\label{deg2}\eeq
where $\alpha_n \in \widehat{U(1)}$ and $L_n \in {\rm Vir}$.
On the degenerate highest-weight state $\vert \{r+n,r\}\rangle_W$,
$\ n\in {\bf Z}_+$, their eigenvalues are, respectively,
\beq
\left\{\alpha_0\right\} \to {2r+n\over\sqrt{2}}\ , \qquad
\left\{ {\alpha_0^2\over 2} + \sum_{l=1}^\infty \alpha_{-l}\alpha_l \right\}
\to {\left(2r+n\right)^2\over 4}\ ,\quad
\left\{ L_0\right\} \to {n^2\over 4}\ .
\label{degei}\eeq
The $\widehat{U(1)}$ mode can be identified with the diagonal
$\u1\subset\u1^{\otimes 2}$, generated by $(J^1 +J^2)/\sqrt{2}$.
The orthogonal current $(J^1 -J^2)/\sqrt{2}$ does not exist in
$\winf$, but gives the Virasoro $L_n$ in (\ref{deg2}) by the
Sugawara construction (\ref{ln}).
The higher $\winf $ generators $V^i_n$ for $i\ge c=2$ are functions of
the operators $\alpha_n $ and $L_n$, and satisfy the rest of the $\winf$
algebra without further conditions \cite{kac2}.
The states of these $\winf$ \reps are built by polynomials of
$\alpha_n$ and $L_n$, with $n<0$; thus, their character is the product
of a $\u1$ character and a $c=1$ Virasoro one.
Actually, the Virasoro representations with dimension $h=n^2/4$
are degenerate, and correspond to those of the
well-known Virasoro $c\le 1$ minimal models \cite{bpz}. They have a
single null vector at level $(n+1)$, and their character is,
\beq
\chi_{M({\rm Vir},1, \{ n^2/4 \} )} =
\eta^{-1}\ q^{n^2/4} \left(1-q^{n+1}\right)\ .
\label{chiv}\eeq
The product of this character and the $\u1$ character in eq.
(\ref{chir}) matches the general formula of the $\winf$ character
(\ref{chid}) for $m=2$ and $\vec{r}=\{r+n,r\}$.
This confirms the one-to-one relation in (\ref{rep2}), and
shows that the $\winf$ null vectors at $c=2$ are those of the Virasoro
algebra at $c=1$.

Moreover, the decomposition of $\u1^{\otimes 2}$ \reps into $\winf$
degenerate \reps can be inferred from the previous characters:
the Virasoro character (\ref{chiv}) can be written as,
\beq
\chi_{M \left({\rm Vir},1, \{ n^2/ 4 \} \right)} =
\chi_{M \left(\u1 ,1, \{n/\sqrt{2} \} \right)} -
\chi_{M \left(\u1 ,1, \{(n+2)/\sqrt{2} \} \right)} \ ,
\label{chivu}\eeq
and solved for the Abelian character. This gives the decomposition
\beq
M \left(\u1^{\otimes 2},2,\{r+n,r\} \right) =\sum_{\ell=0}^\infty
M \left(\winf, 2, \{ r+n+\ell,r-\ell \} \right)\ .
\label{deco}\eeq

The fusion rules of the degenerate $\winf$ representations are different
from the Abelian addition of weights (\ref{frule}), due to the non-trivial
structure of excited states. For $c=2$, we can deduce them from the
fusion rules of the Virasoro minimal models for $c\to 1$ \cite{bpz}.
By denoting $M({\rm Vir},1, \{ n^2/ 4 \} )\equiv [n]$, the latter are
\beq
{[n]}\bullet [k] = [n+k] +[n+k-2] + \cdots +[|n-k|]\ ,
\label{fusvir}\eeq
and actually correspond to the addition of two $SU(2)$ isospin
quantum numbers, identified as $s_1=n/2$ and $s_2=k/2$, respectively.
{}From the previous construction of $\winf\sim\u1\otimes{\rm Vir}$,
we deduce the fusion rules
\barr
&&M\left(\winf,2,\vec{r}\right) \bullet M\left(\winf,2,\vec{s}\right)
  =\sum_{\displaystyle{\vec{t}\in\Lambda_{\vec{r},\vec{s}}}}
  M\left(\winf,2,\vec{t}\right)\ ,\nonumber\\
&&\Lambda_{\vec{r},\vec{s}}=\left\{ \vec{t} \left\vert
  \vec{t}=\vec{r}+\vec{s}-\ell \vec{\alpha} \right. ,
  \ 0\le \ell\le {n+m-|n-m|\over 2} \right\}\ ,\nonumber\\
&&\vec{r}=\left({r+n \atop r} \right)\ ,\
  \vec{s}=\left({s+m \atop s} \right)\ ,\
  \vec{\alpha}=\left({+ 1 \atop -1} \right)\ ,
  \ n,m,\ell\in {\bf Z}_+ \ .
\label{latw2}\earr

A general fact in conformal field theory is that the fusion of
degenerate \reps only gives degenerate \reps of the same type; thus
it is possible to find sets of degenerate \reps which are closed
under the fusion rules \cite{bpz}.

\bigskip

\noindent{\bf Construction of the minimal models}

The $\winf$ minimal models are constructed by assembling the minimal
set of one-class degenerate representations (\ref{chid}) which is closed
under fusion.
This algebraic method of construction, widely applied in conformal field
theory, was already introduced for the Abelian model of section 2.
In that case, the lattice $\Gamma$ (\ref{latt}) was shown to be the minimal
set of $\u1^{\otimes m}$ weights which is closed under the addition as
fusion rule (\ref{frule}).
The fusion rule of $c=2$ degenerate $\winf$ weights (\ref{latw2}) is
again the addition, but modulo the special weight
$\vec{\alpha}=\{1,-1\}$ (with vanishing $\u1$ charge).
As a consequence, these rules close again on a lattice ${\rm P}^+$
which satisfies some conditions:
i)\ the points in ${\rm P}^+$, in particular the basis vectors,
should all be degenerate weights;
ii)\ the weight $\vec{\alpha}$ should be in ${\rm P}^+$, such that the
result of the fusion (\ref{latw2}) is a sum of weights in ${\rm P}^+$,
which belongs to ${\rm P}^+$;
iii)\ the ordering of weight components in (\ref{chid}), $r_1\ge r_2$,
being preserved by fusion, should be imposed to avoid double counting.
The following basis,
\beq
\vec{v}_1= {s+k \choose s}\ ,\quad \vec{v}_2 ={+ 1 \choose -1}\ ,
\quad s\in {\rm R}\ ,\ k\in {\rm Z\ \ odd}\ ,
\label{bas2}\eeq
conveniently expresses these properties of ${\rm P}^+$,
\beq
{\rm P}^+=\left\{ \vec{r} \left\vert \vec{r}=n_1\vec{v}_1 +n_2 \vec{v}_2\ ,
n_1,n_2 \in {\bf Z}\ ,\ kn_1 +n_2 \ge 0 \right.\right\}
\qquad (c=2)\ .
\label{pplus2}\eeq
Note that $k$ is taken odd in eq.(\ref{bas2}) in order to span both integer
and half-integer isospin representations.

Next, we extend these results to $m>2$ by using some results of the
${\cal W}_m$ minimal models at $c=(m-1)$ \cite{fz} :

i) The ${\cal W}_m$ degenerate representations are in one-to-one relation
with the \reps of the $SU(m)$ Lie algebra, which are characterized
by a $(m-1)$ dimensional highest-weight vector ${\bf \Lambda}$ ,
\beq
{\bf\Lambda} = \sum_{a=1}^{m-1}\ {\bf\Lambda}^{(a)}\ \ell_a\ ,\qquad
\ell_a \in {\bf Z}\ ,
\label{weigm}\eeq
where ${\bf\Lambda}^{(a)}$ are the fundamental weights of $SU(m)$ \cite{wy}.
In particular, the ${\cal W}_m$ fusion rules are isomorphic to the
decomposition of $SU(m)$ tensor representations.

ii) The eigenvalue $h$ of the Virasoro generator $L_0$ for these \reps is
given by
\beq
2h =\left\Vert \sum_{a=1}^{m-1}\ {\bf\Lambda}^{(a)}\ \ell_a\right\Vert^2\ ,
\label{virm}\eeq
i.e. by the norm of the highest weight.
Due to the equivalence $\winf\sim\u1 \otimes {\cal W}_m$,
the $m$-dimensional $\winf$ weight $\vec{r}$ can be written in terms of
the ${\cal W}_m$ weight.
To this extent, we introduce the new basis $\vec{q}=\{q,{\bf\Lambda}\}$
for the $\winf$ weights by the orthogonal transformation
$\ q_i = \sum_{j=1}^m\  U_{ij}\ r_j \ $ , defined by,
\barr
q &=& {1\over\sqrt{m}} \ \left( r_1+r_2+\cdots +r_m \right)\ ,\nonumber\\
\Lambda_a &=& \sum_{i=1}^m\ u^{(i)}_a \ r_i\ ,\qquad a=1,\dots, m-1\ ,
\label{utran}
\earr
where the $m$, $(m-1)$-dimensional vectors ${\bf u}^{(i)}$ are
\beq
\begin{array}{l c r}
  {\bf u}^{(1)} = \left(
  \ \ {1\over\sqrt{2}}\ ,\ {1\over\sqrt{6}}\ ,\ {1\over\sqrt{12}}\ ,\dots,
  \right.
  &{1\over\sqrt{k(k+1)}}\ ,\dots, &\left.{1\over\sqrt{m(m-1)}}\right)\ ,
\\
  {\bf u}^{(2)} = \left(
  - {1\over\sqrt{2}}\ ,\ {1\over\sqrt{6}}\ , \dots\dots\dots\right.
  &\dots\dots\dots\dots, &\left. {1\over\sqrt{m(m-1)}} \right)\ ,
\\
  {\bf u}^{(k)} = \left(
  \overbrace{\ \ 0\ ,\dots\dots\dots\dots,\ 0\ ,}^{k-2} \right.
  &- {k-1\over\sqrt{k(k-1)}}\ ,\dots,&\left. {1\over\sqrt{m(m-1)}}\right)\ ,
\\
  {\bf u}^{(m)} = \left(
  \ \ 0\ , \dots\dots\dots\dots\dots \right.
  &\dots\dots\dots\dots ,\ 0\ , &\left. -{m-1\over\sqrt{m(m-1)}} \right)\ .
\end{array}
\label{uvect}\eeq

These vectors satisfy the rules
\beq
\sum_{i=1}^m\ {\bf u}^{(i)} =0\ ,\qquad
\sum_{i=1}^m\ u^{(i)}_a\ u^{(i)}_b =\delta_{ab}\ ,\qquad
{\bf u}^{(i)}\cdot {\bf u}^{(j)} = \delta_{ij} -{1\over m}\ ,
\label{urules}\eeq
and can be used to build the lattice of roots and weights of $SU(m)$ \cite{wy}.
The simple roots ${\bf\alpha}^{(a)}$ and the fundamental weights
${\bf \Lambda}^{(a)}$ are
\barr
&&{\bf\alpha}^{(a)}= {\bf u}^{(a)} -{\bf u}^{(a+1)}\ ,\nonumber\\
&&{\bf\Lambda}^{(a)} = \sum_{i=1}^a\ {\bf u}^{(i)}\ ,
\qquad a=1,\dots, m-1\ , \nonumber\\
&&{\bf\Lambda}^{(a)} \cdot {\bf\alpha}^{(b)} = \delta_{ab}\ ,\qquad
{\bf\Lambda}^{(a)}\cdot {\bf\Lambda}^{(b)} = a- {ab\over m}\ ,
\quad a \le b \ .
\label{rootwei}\earr
After these definitions, it easy to see that the equations (\ref{weigm}) and
(\ref{utran}) are made equal by identifying the integer components of the
$\winf$ degenerate weight $(r_a-r_{a+1})$ with those of the $SU(m)$ weight
$\ell_a\ $:
\beq
\ell_a =r_a -r_{a+1}\ , \qquad a=1,\dots,m-1\ .
\label{sumid}\eeq
It is useful to check the $m=2$ case in this $SU(m)$ notation:
\barr
u^{(1)}&=&{1\over\sqrt{2}}\ ,\quad u^{(2)}=-{1\over\sqrt{2}}\ ,\quad
\alpha^{(1)} = \sqrt{2}\ ,\quad \Lambda^{(1)}= {1\over\sqrt{2}}\ ,\nonumber\\
q&=&{1\over\sqrt{2}}\left( r_1+r_2 \right)\ ,\nonumber\\
\Lambda &=& {1\over\sqrt{2}} \left( r_1-r_2\right) =
\ell_1\ \Lambda^{(1)}\ , \quad \ell_1\in {\bf Z}_+\ .
\label{m2case}\earr
In the $\{q,\Lambda\}$ basis, the $m=2$ fusion rules (\ref{latw2}) are
\beq
\vec{q} \bullet \vec{p} = \vec{p}+\vec{q} \ ,\qquad {\rm mod}\quad
{0\choose \sqrt{2} } = {0 \choose \alpha^{(1)} } \ ,\quad (m=2)\ .
\label{fus22}\eeq
Based on these data, we can identify the components of the $\winf$
weight in the basis $\vec{q}=\{q,{\bf \Lambda}\}$ (\ref{utran}) as the
$\u1$ charge and the $SU(m)$ weight, respectively, which label
the tensor representations $\u1\otimes {\cal W}_m$.
We can thus deduce the $\winf$ fusion rules from the $SU(m)$ fusion rules,
and find their general structure:
\beq
\vec{q} \bullet \vec{p} = \vec{p}+\vec{q} \ ,\qquad
{\rm mod}\quad \left\{ {0 \choose {\bf\alpha}^{(1)} }\ ,\cdots,
 {0 \choose {\bf\alpha}^{(m-1)} } \right\} \ ,
\label{fusm}\eeq
namely the $SU(m)$ weights add up modulo an integer combination
of the simple roots \cite{wy}.

Similarly to the $(m=2)$ case, we can find lattices ${\rm P}^+$ which
are closed under these fusion rules because they contain the
$(m-1)$ simple roots. In the original $\vec{r}$ basis (\ref{chid}),
these lattices can be generated as follows
(see eqs.(\ref{rootwei},\ref{chid})),
\beq
\vec{r}=\sum_{i=1}^m n_i\ \vec{v}_i=
n_1\left(\begin{array}{c}
s+k_1 \\ .\\ .\\ .\\ s+k_{m-1}\\ s\end{array}\right)+
n_2\left(\begin{array}{c} +1 \\ -1 \\ 0 \\ .\\ .\\ 0 \end{array}\right)+
n_3\left(\begin{array}{c} 0 \\ +1 \\ -1 \\ 0 \\ . \\ 0 \end{array}\right)+
\cdots +
n_m\left(\begin{array}{c} 0 \\ . \\ . \\ 0\\ +1 \\ -1 \end{array}\right)\ ,
\label{latm}\eeq
with $n_i\ ,\ k_j \in {\bf Z}$. The condition
\beq
\xi  \equiv -{(-1)^m+k_1+\cdots+k_{m-1} \over m} \in {\bf Z}\ ,
\label{nality}\eeq
ensures that $SU(m)$ \reps of minimal n-ality, e.g. the fundamental
one, are contained in ${\rm P}^+$.
The ordering of components $r_a \ge r_{a+1}$ will be discussed later on.

In summary, the $\winf$ minimal models are made of one-class
degenerate \reps (\ref{chid}) with weights belonging to the lattices
(\ref{latm}), which are parametrized by $s\in{\rm R}$ and $\{k_j\}\in {\bf Z}$,
$j=1,\dots,m-1$, with the constraints (\ref{nality}) and $r_a \ge r_{a+1}$,
$\ a=1,\dots,m-1$.

Further conditions on the parameters of these lattices, i.e. on
the $\winf$ minimal models, come from a physical requirement already
discussed in the $\u1^{\otimes m}$ theory.
We must identify the electron excitation in the lattice (\ref{latm})
and define the corresponding unit of charge.
It is convenient to introduce another simpler basis $\vec{u}_i$
for the lattice (\ref{nality}), by the following modular
transformation (\ref{modinv}) $\vec{v}_i=\sum_{j=1}^m\ \Xi_{ij}\ \vec{u}_j$,
\beq
\Xi = \left( \begin{array}{c c c c c c}
\xi  +k_1 & \xi  +k_2 & \cdots & \cdots & \xi  +k_{m-1} & \xi   \\
1     & -1    & 0      & \cdots & \cdots    & 0 \\
0     & 1     & -1     & \cdots & \cdots    & 0 \\
\cdots&\cdots & \cdots & \cdots & \cdots    & \cdots\\
0    & \cdots & \cdots & \cdots & 1         & -1
\end{array} \right)\ \in SL(m,{\bf Z}) \ .
\eeq
This leads to the vectors
\beq
\left( \vec{u}_i\right)_j =\delta_{ij} + rC_{ij} \ ,
\qquad r =(-1)^m(\xi  -s) \in {\rm R}\ .
\label{modtrans}\eeq

Given that the $\winf$ and the $\u1^{\otimes m}$ \reps are
labelled by the same type of highest-weight vector $\vec{r}$,
the spectrum of the corresponding quantum numbers $(\nu,Q,\theta/\pi)$
of the $\winf$ minimal models can be again described by the
$K$ matrix (\ref{qtheta},\ref{sig}), which is the metric of the lattice
${\rm P}^+$ in the basis (\ref{modtrans}):
\beq
K^{-1}_{ij} = \vec{u}_i \cdot \vec{u}_j =\delta_{ij} +\lambda C_{ij} \ ,
\quad \lambda =mr^2 + 2r \in {\rm R}\ ,
\label{kmin}\eeq
These quantum numbers can be written (see eq.(\ref{qtheta})),
\barr
\nu & =& t^T \cdot K^{-1} \cdot t \ ,\nonumber\\
Q & = & t^T \cdot K^{-1} \cdot n \ , \qquad n_i \in {\bf Z}\ ,
\qquad n_1 \ge n_2 \ge\dots\ge n_m\ , \nonumber\\
{\theta\over\pi} & = & n^T \cdot K^{-1}\cdot n \ .
\label{minspec}\earr
In this equation, the arbitrary vector $t^T=(1,\dots,1)\cdot \Theta$ ,
 $\Theta\in Sl(m,{\bf Z})$, parametrizes any possible choice of basis for the
lattice ${\rm P}^+$ (\ref{latm}). This vector defines the physical charge $Q$
as a linear functional of the $\winf$ weights, and is determined as follows.
The $\winf$ degenerate \reps, being equivalent to the
$\u1\otimes {\cal W}_m$ ones, are characterized by a unique additive quantum
number $q$ in (\ref{utran}), which can be interpreted as the electric charge.
Actually, the $\winf$ minimal models do not possess the notion of
$m$ individual edges as in the Abelian hierarchical theory (\ref{bosact}),
corresponding to its $m$ currents.
Therefore, the physical total charge $Q$ in (\ref{minspec}) should be taken
proportional to $q$.
By working through the previous coordinate changes, this condition
can be translated into the condition
$\vec{t} \propto \{1,\dots, 1\}$ , whose unique solution is
\beq
\vec{t}=\{1,\dots,1\}\ ,
\label{qfix}\eeq
because all $\Theta$ have unit determinant.
We have thus obtained $K$ matrices of the $\winf$ minimal models (\ref{kmin})
which are almost equal to those of the Jain hierarchy (\ref{kjain}); only the
quantization of the parameter $\lambda$ is to be found and the ordering
$n_a \ge n_{a+1}$ to be explained.

To this end, we require that the spectrum (\ref{minspec}), with
$\vec{t}=\{1,\dots,1\}$, contains {\it one} excitation $n_i=e_i \in {\bf Z}$,
$i=1,\dots,m$, with the quantum numbers and monodromy properties of the
electron.
The monodromy properties of a pair of excitations are dictated by the fusion
rules of the corresponding $\winf$ degenerate \reps \cite{bpz}.
In the monodromy process of the pair $(n_i,\ m_i)$, more than one
intermediate state $k_i$ is obtained by fusion: each one
acquires a phase proportional to the Virasoro eigenvalues, which is
\hbox{ $\theta[n_i] +\theta[m_i]-\theta[k_i]$},
where $\theta[n_i]$ is defined in (\ref{minspec}).
This is the {\it non-Abelian} fractional
statistics which will be better analyzed in the next section.
It can be shown that the Virasoro eigenvalues of the intermediate states only
differ by integers, thus the locality conditions for the electron excitations
reduce to those for the Abelian fusion rules ($k_i=n_i+m_i$).
These conditions are:
\barr
&(i)&\quad Q\left[n_i=e_i\right] = 1\ ,
\qquad (ii) \qquad {\theta [e_i] \over\pi} \in {\rm Z\ odd}\ ,\nonumber\\
&(iii)& \quad {1\over 2\pi}\left(
\theta[e_i] +\theta[n_i]-\theta[e_i+n_i] \right)
 = \ e^T \cdot K^{-1}\cdot n \ \in {\bf Z}\ \ \forall\ n_i\ .
\label{Qdef}\earr
The condition $(iii)$ requires that
$\sum_{j=1}^m\ K^{-1}_{ij}\ e_j=\eta_i\in{\bf Z}\ \ \forall\ i$,
and the condition $(i)$ that $\sum_i\eta_i =1$.
Note that the $K$ matrices (\ref{kjain}) are invariant under permutations
of the basis components,
\beq
K_{ij} =K_{\sigma(i)\sigma(j)}\ , \qquad \sigma\in {\cal S}_m\ ,
\label{perm}\eeq
where ${\cal S}_m$ is the symmetric group. Due to this symmetry,
we can choose the solution of $(i)$ in the form
$\eta_i=(1,0,0,\dots,0)$, and obtain the electron excitation as
\beq
e_i=(1+p,p,\dots,p) \ ,
\label{espec}\eeq
where $p=-\lambda/(1+m\lambda) \in {\bf Z}$.
Moreover, the condition $(ii)$ requires $e\cdot \eta =1+p\ $ odd, i.e.
$p$ even.

We have thus obtained the Jain $K$ matrices $K_{ij} =\delta_{ij} +p\ C_{ij}$,
in the purely chiral case ($p>0$ even). These are positive definite by
construction (eq.(\ref{kmin})).
Mixed chiral and antichiral propagation of the edge excitations can be
introduced by extending the arguments used in the discussion of the
chiral boson hierarchy of sect. 2 (see eqs.(\ref{schwi}-\ref{achi}).
The data of the $\winf$ \reps, summarized in the lattice $\Gamma$
(\ref{latm}), are independent of the choice of chirality -
only the identification of the physical quantities changes.
For chiral (antichiral) excitations, we must identify the fractional
statistics spectrum $\theta/\pi$ with plus (minus) the Virasoro eigenvalue,
while the charge eigenvalue does not change.
In the $\winf$ minimal models, made by $\u1\times {\cal W}_m$
representations, two types
of excitations can be identified: the charged excitations associated to the
$\u1$ factor and the neutral, non-standard, excitations described by the
${\cal W}_m$ factor.
Therefore, there are only two possible choices, corresponding
to whether the chiralities of charged and neutral excitations are equal or
opposite. The $(m-1)$ neutral excitations described by ${\cal W}_m$ are
{\it interacting } and cannot be assigned mixed chiralities.

In order to change the sign of the corresponding Virasoro eigenvalues, we
diagonalize the quadratic form of $\theta/\pi$ in (\ref{minspec}) by the
orthogonal transformation $K=U^T D U$, where $U$ was defined
in (\ref{utran}). We can thus summarize the relations among the integer
labels $\{\vec{n}\}$ in eq.(\ref{minspec}), the $\winf$ weights
$\{\vec{r}\}$ in eq.(\ref{chid}) and the $\u1\otimes {\cal W}_m$ weights
$\{\vec{q}=(q,{\bf \Lambda})\}$ in eq.(\ref{utran}), as follows:
\barr
\vec{r} & =& U^T D^{-1/2} U \cdot \vec{n}\ ,\nonumber\\
\vec{q} & =& D^{-1/2} U \cdot \vec{n} = U \cdot \vec{r} \ ,\nonumber\\
K &= &U^T D U \ ,\qquad
D^{1/2}\equiv\ {\rm diag}\left( \sqrt{1+mp}, 1, \dots, 1\right)\ ,\quad p>0\ .
\label{bmess}\earr
In components,
\barr
q &=& {1\over \sqrt{m}} \sum_{i=1}^m\ r_i\ =\ {1\over \sqrt{m}}
{1\over \sqrt{1+mp}} \sum_{i=1}^m\ n_i\ ,\nonumber\\
{\bf\Lambda} &=& \sum_{i=1}^m {\bf u}^{(i)}\ r_i =
\sum_{i=1}^m {\bf u}^{(i)}\ n_i \ , \nonumber\\
i.e. &\ & r_a -r_{a+1} = n_a -n_{a+1}\ , \qquad a=1,\dots,m-1\ .
\label{bmesss}\earr
Therefore, in the chiral case, the diagonal form of the Virasoro eigenvalue is,
\beq
{\theta\over\pi}= 2h= n^T\cdot K^{-1}\cdot n= \sum_{i=1}^m r_i^2=q^2 +
\left\Vert {\bf\Lambda} \right\Vert^2 \ .
\label{diagv}\eeq
We can change the relative chirality of the neutral and charged excitations
by defining
\beq
{\widetilde\theta\over\pi}=q^2 - \left\Vert {\bf\Lambda} \right\Vert^2 =
n^T\cdot \widetilde{K}^{-1} \cdot n \ .
\label{chianti}\eeq
The solution of the discretization electron conditions now gives \\
$\widetilde{D}={\rm diag}\ \left( (mp-1), -1, \dots, -1\right)$, i.e.
$\widetilde{K}=p\ C_{ij} -\delta_{ij}$, with $p$ even positive integer,
which we recognize as the Jain matrices for mixed chiralities (\ref{kjain}).
Finally, the ordering of $\vec{r}$ components in (\ref{chid}) translates into
the announced conditions $n_a \ge n_{a+1}$, which identify the minimal lattice
${\rm P}^+$ (\ref{latm}).

%\vfill\eject

%-4---------------
\vbox to .5in {\vfill}
\section{Physical properties of $\winf$ minimal models}

In the previous section, we have constructed the $\winf$ minimal models
out of one-congruence-class degenerate \reps (\ref{chid}) \cite{kac2}.
We have found their filling fractions and the spectrum of
fractional charge and statistics of their edge excitations
(\ref{minspec},\ref{bmess}),
\barr
\nu& =& {m\over mp \pm 1}\ , \qquad p >0 \ {\rm even}\ ,\qquad c = m\ ,
\nonumber\\
Q & = & {1\over pm \pm 1}\ \sum_{i=1}^m n_i \ ,\qquad
n_1 \ge n_2 \ge \cdots \ge n_m\ , \nonumber\\
{ \theta\over\pi} &=& \pm\left( \sum_{i=1}^m n^2_i -
{p\over mp \pm 1}\left( \sum_{i=1}^m n_i \right)^2 \right) \ .
\label{wspec}\earr
These spectra agree with the experimental data and match the results
of the lowest-order Jain hierarchy (\ref{jspec})\footnote{
Note, however, the reduced multiplicities of eq.(\ref{wspec}).}
discussed in section 2.
Our derivation was independent and self-consistent within the theory of
edge excitations; we only used the principle of $\winf$ symmetry of the
incompressible fluids and the hypothesis of stability of the
minimal theories, and we did not make any reference to microscopic
physics and wave-functions.
This independent hierarchical construction is the main result of this
paper.

Furthermore, the detailed predictions of the $\winf$ minimal theories
are different from those of the chiral boson theories.
The neutral excitations are associated to ${\cal W}_m$ \reps and carry an
$SU(m)$ quantum number, leading to new physical effects.
We shall discuss two of them:

i) The non-Abelian fusion rules and non-Abelian fractional statistics;

ii) The degeneracy of excitations at fixed angular momentum
above the ground state (the Wen topological order on the
disk geometry \cite{topord}).

\bigskip

\noindent{\bf Non-Abelian fusion rules and non-Abelian statistics}

In the previous section, we have identified the physical electron
as the minimal set of $\winf$ \reps with unit charge and
integer statistics relative to all excitations. These conditions are
fulfilled by a composite edge excitation $n_i=(1+p,p,\dots,p)$,
which is made of $(mp)$ elementary charged {\it anyons} and the
{\it quark} elementary neutral excitation, i.e. the fundamental
$SU(m)$ isospin \rep, due to $(n_i-n_{i+1})=\delta_{i,1}$.

A conduction experiment which could show the composite nature of
the electron has been proposed \cite{kane}.
It is a modification of the ``time-domain'' experiment \cite{tdom},
in which a very fast electric pulse was injected at the boundary
of a disk sample and a chiral wave was detected at another
boundary point. The proposed experiment will also detect the neutral
excitation in the electron, which propagates at a different speed.

The compositeness of the electron also plays a role in
the resonant tunnelling experiment \cite{milli}, in which
two edges of the sample are pinched at one point, such that the
corresponding edge excitations, having opposite chiralities, can interact.
At $\nu=1/3$, the point interaction of two elementary anyons
is relevant and determines the {\it scaling law} $T^{2/3}$ for
the conductance \cite{tunn}.
This scaling of the tunnelling resonance peaks is verified experimentally.
On the other hand, off-resonance and at low temperature, the conductance
is given by the tunnelling of the whole electron, with a different
scaling law in temperature \cite{wen}.

These experiments involve processes with one or two electrons:
their quark compositeness can be seen in four-electron processes,
like scattering.
Indeed, the expansion of the four-point function of the
electrons in intermediate channels is determined by the
fusion ((\ref{fusvir}) for $m=2$) of an electron pair. This is, schematically,
\beq
\langle\Omega| \Psi^\dagger(1)\Psi^\dagger(2)\Psi(3)\Psi(4) |\Omega\rangle
=\sum_{s=0,1} \langle\Omega| \Psi^\dagger(1)\Psi^\dagger(2) |\{s\}\rangle
\langle \{s\}| \Psi(3)\Psi(4)|\Omega\rangle\ ,
\label{nonab}\eeq
where the two channels follow from the addition of two one-half isospin values.
More than one intermediate channel are also created in the
adiabatic transport of two electrons around each other, in presence of
two other excitations, because the amplitude for this process is again a
four-point function. For generic excitations, the monodromy phases form a
matrix, which gives a non-Abelian
representation of the braid group \cite{wilc}.
This is precisely the notion of non-Abelian statistics\footnote{
For a general discussion of non-Abelian statistics in the quantum Hall
effect, see ref.\cite{moore}.}.
These monodromy properties also determine the degeneracy of
the ground state on a torus geometry, the so-called topological order
\cite{wen}. This depends on the type of the \reps carried
by the edge excitations \cite{bpz}, and should be computed for the
$\u1\otimes{\cal W}_m$ ones.
We hope to develop these issues in a separate work.

\bigskip

\noindent{\bf The degeneracy of excitations above the ground state}

In order to discuss this point, we must rewrite the spectrum (\ref{wspec}).
Let us consider the $m=2$ chiral theories, relevant for
$\nu=2/5, \dots$; the extension of the analysis to any $m$ and
mixed chiralities is straightforward.
Recall that any excitation $(n_1,n_2)$ is associated to a $\u1\otimes{\rm Vir}$
\rep, labelled by the $\u1$ charge $Q\propto (n_1+n_2)$ and the
$SU(2)$ isospin $s=|n_1-n_2|/2$ (\ref{fusvir}).
Divide the square lattice $(n_1,n_2)$
into charged excitations and their neutral daughter excitations by introducing
the change of integer variables $(n_1,n_2) \to (l, n)$:
\beq
{\rm I}:\left\{
\begin{array}{l l}
2 l & =n_1+n_2 \\ 2n & =n_1-n_2 > 0 \\
& \ \ (n_1+n_2 \ {\rm even}), \end{array} \right.
\qquad\quad
{\rm II}:\left\{
\begin{array}{l l}
2 l +1& =n_1+n_2 \\ 2n +1& =n_1-n_2 > 0 \\
& \ \ (n_1+n_2 \ {\rm odd}). \end{array} \right.
\label{intcha}\eeq
The spectrum (\ref{wspec}) can be rewritten, for $\nu=2/(2p+1)$,
\beq
{\rm I}:\left\{
\begin{array}{l l}
Q & = {2l \over 2p+1}\ , \\
{1\over 2}{\theta\over \pi} & = {1\over 2p+1} l^2 +n^2  \end{array} \right.
\qquad\qquad
{\rm II}:\left\{
\begin{array}{l l}
Q & = {2 \over 2p+1}\left( l +{1\over 2} \right) \\
{1\over 2}{\theta\over \pi} & =
{1\over 2p+1} \left( l+{1\over 2} \right)^2 +{(2n+1)^2 \over 4}
\end{array} \right. \ .
\label{splitsp}\eeq
The $\u1$ charged excitations have  the same spectrum
$Q=\nu k,\ \theta/\pi=\nu k^2 $, of the simpler Laughlin fluids $(m=1)$.
Moreover, the infinite tower of neutral daughters ($n>0$) are characterized
by the conformal dimensions $h=(2n)^2/4\ $ (resp. $h=(2n+1)^2/4$).

The number of excitations above the ground state depends
on whether the neutral excitations have a bulk gap or not.
This affects also the thermodynamic quantities like the specific heat.

As said before, the charged edge excitations correspond to Laughlin
quasi-particles vortices in the bulk of the incompressible fluid,
which spill their density excess or defect to the boundary.
They have an (non-universal) gap proportional to the electrostatic energy of
the vortex core, which is not accounted for by the edge theory \cite{laugh}.
On the other hand, the bulk excitations corresponding to {\it neutral} edge
excitations are not well understood yet.
If they have a gap, they could exibit the internal structure of the
quasi-particle vortex, or be bound states of a quasi-particle and a
quasi-hole; these would be localized two-dimensional excitations.
Neutral and charged gapful excitations can be thought of as analogs of the
{\it breathers} and {\it solitons} of one-dimensional integrable models,
respectively, \cite{inte}.
On the other hand, gapless neutral excitations would be pure effects
of the structured edge.

In the gapful case, the excitations above the ground state are
particle-hole excitations described by the $\winf$ \rep of the ground state
$(n=l=0)$ in (\ref{splitsp}).
In the gapless case, there are also contributions from the neutral
daughter Virasoro representations ($n>0\ ,\ l=0$), because they have
integer spin (Virasoro dimension) and are indistinguishable.
Actually, the infinite tower of Virasoro \reps
($n>0\ ,\ l$ fixed) of each charged parent state $(l,n=0)$
can be summed (with their multiplicity one) into a single $\u1$ \rep
by using backward the decomposition (\ref{deco}):
\barr
{\rm I} & : & \sum_{n=0}^\infty\ M \left( {\rm Vir},1, \{ n^2 \} \right) =
M \left( \u1 ,1, \{ 0 \} \right) \ ,\nonumber\\
{\rm II} & : & \sum_{n=0}^\infty\
M \left( {\rm Vir},1, \{ (2n+1)^2/ 4 \} \right) =
M \left( \u1 ,1, \{ 1/\sqrt{2} \} \right) \ .
\label{deconstr}\earr
In this case, the $m=2$ $\winf$ square-lattice spectrum (\ref{splitsp})
reduces to a one-dimensional array of $\u1^{\otimes 2}$ \reps,
with spectrum
\beq
{\rm I}: \ {1\over 2}{\theta\over\pi} ={1\over 2p+1} l^2\ ,\qquad\qquad
{\rm II}: \ {1\over 2}{\theta\over\pi} =
{1\over 2p+1} \left(l+{1\over 2} \right)^2 + {1\over 4}\ ,
\label{triv}\eeq
where the second $\u1$ eigenvalue is not observable.

Let us repeat this analysis for the corresponding chiral boson theory
of the Jain fluid. The spectrum of charge and fractional statistics is
again given by (\ref{splitsp}), with multiplicities given by
$n\in {\rm Z}$: each $(l,n)$ value corresponds to a $\u1\otimes\u1$
\rep now.
If they are gapless, the neutral daughter
$\u1\otimes\u1$ representations $((l,n),\ n\neq 0\in {\bf Z})$ of
each charged \rep $(l,0)$ can be similarly summed up into
one \rep of the larger algebra $\u1\otimes\widehat{SU(2)}_1$,
the non-Abelian current algebra of level one \cite{bpz}.
Using the following expansion of the $\widehat{SU(2)}_1$ characters
of spin $s=0$ and $s=1/2$ \cite{bpz}\cite{ciz} :
\barr
I & : & \chi_{M(\widehat{SU(2)}_1,1,s=0)}= \chi_{M(\u1,1,r=0)} +
     2 \sum_{n=1}^\infty\ \chi_{M(\u1,1,r=\sqrt{2}n)}\ ,\nonumber\\
II & : & \chi_{M(\widehat{SU(2)}_1,1,s=1/2)}=
     2 \sum_{n=0}^\infty\ \chi_{M(\u1,1,r=(2n+1)/\sqrt{2})}\ ,
\label{su2char}\earr
we sum all neutral states $(n \in {\rm Z})$ with the correct multiplicity.
The spectrum of $\u1\otimes\widehat{SU(2)}_1$ representations is again
given by (\ref{triv}).

We can now compare the predictions of the $\winf$ minimal models and
the chiral boson theories for the degeneracy of the excitations above the
ground state.
This degeneracy can be measured in numerical simulations of a few electron
system in the disk geometry, by charting the eigenstates
of the Hamiltonian below the bulk gap \cite{wen}\cite{topord}.
Consider, for example, the $\nu=2/5$
$\ (m=p=2)\ $ ground state ($(l=0,n=0)$ in (\ref{splitsp})).
In the following table, we report the degeneracies encoded in the
$\u1\otimes {\rm Vir}$ character (\ref{chid})
and the $\u1^{\otimes 2}$ character (\ref{chir}), for $\vec{r}=0$,
as well as those of the $\u1\otimes\widehat{SU(2)}_1$ one (\ref{su2char})
for $r=s=0$:
\beq
\begin{array}{c | r r r r r r}
 \Delta J                   & 0 & 1 & 2 & 3  & 4  & 5  \\ \hline
 \u1\otimes {\rm Vir}       & 1 & 1 & 3 & 5  & 10 & 16  \\
 \u1\otimes\u1              & 1 & 2 & 5 & 10 & 20 & 36  \\
 \u1\otimes\widehat{SU(2)}_1& 1 & 4 & 9 & 20 & 42 & 80
\end{array}
\label{degtab}\eeq

If neutral daughter excitations have a gap, they should
not be counted, and the degeneracy is only given by the particle-hole
excitations encoded in the ground state character of the theory.
On the other hand, gapless neutral excitations contribute and the total
degeneracy is given by the resummed characters (\ref{deconstr},\ref{su2char}).
We conclude that:

i) The observation of $\u1\otimes {\rm Vir}$ degeneracies confirms
the $\winf$ minimal theory with gapful neutral excitations;

ii) The $\u1\otimes\u1$ degeneracies are found both in the $\winf$ minimal
theory with gapless neutral excitations and in the chiral boson theory with
gapful ones;

iii) The $\u1\otimes\widehat{SU(2)}_1$ degeneracies support the chiral boson
theory with gapless neutral excitations.

Numerical results known to us \cite{wen} are not accurate
enough to see the differences in table (\ref{degtab}).
Note the characteristic reduction of states of $\winf$ minimal models.

These remarks on the gap for neutral excitations do not affect
the previous discussion of the conduction experiments,
where excitations move along one edge or are transferred between two edges
at the same Fermi energy, such that bulk excitations are never produced.
Although the resummation of the neutral daughter ${\cal W}_m$ excitations gives
Abelian excitations, these are not $\winf$ irreducible, and thus unlikely to be
produced experimentally.
We think that only irreducible $\winf$ excitations, i.e.,
the elementary ones, can be naturally produced in a real system by an
external probe, for example by injecting an electron at the edge.

\bigskip

\noindent{\bf Remarks on the $SU(m)$ and $\widehat{SU(m)}_1$ symmetries}

We would like to explain the type of non-Abelian symmetry of
the $\winf$ minimal models
and clarify the differences with the chiral boson theories of the Jain
hierarchy, which have been also assigned the $SU(m)$ and $\widehat{SU(m)}_1$
symmetries \cite{read}\cite{kmat}\cite{froh}\cite{kane}.

Due to the $\u1\otimes{\cal W}_m$ construction of the $\winf$ models,
their excitations carry a quantum number which adds up as a $SU(m)$ isospin.
This {\it does not} imply that these models have the full $SU(m)$ symmetry,
in the usual sense of, say, the quark model of strong interactions,
because the states in each ${\cal W}_m$ representation do not form $SU(m)$
multiplets. As shown by the $m=2$ case, the quantum number $s=n/2$ of
Virasoro \reps is like the total isospin $S^2=s(s+1)$, but the $S_z$ component
is missing.
In some sense, the effects of the ${\cal W}_m$ non-Abelian fusion rules can
be thought of as a {\it hidden} $SU(m)$ symmetry.

On the other hand, it has been claimed that the chiral boson theories of the
Jain hierarchy  have a $SU(m)$ symmetry. The correct statement is, however,
that they possess $\u1\otimes\widehat{SU(m)}_1$ symmetry.
This means that their $\u1^{\otimes m}$ representations
can be rearranged into \reps of the $\u1\otimes\widehat{SU(m)}_1$
current algebra, as shown before in (\ref{su2char}).
In the $\widehat{SU(m)}_k$ current algebra, the weights cannot be arbitrary,
but are cut-off by the {\it level} $k$  (e.g. for $m=2$, the spin $s$
can be $0\le s\le k/2$) \cite{bpz}. The level-one non-Abelian
current algebra has very elementary representations and their
fusion rules are made Abelian by this cut-off.

Therefore, the $\widehat{SU(m)}_1$ symmetry has no non-Abelian
physical effect, it is only a convenient reorganization of the
Abelian current algebra.
The non-Abelian character of the excitations is a characteristic
feature of the $\winf$ minimal models.
%%

%\vfill\eject

%-5-------------------

\section{Concluding remarks}

In this paper, we constructed the {\it simplest} $\winf$ minimal models, which
are made of one-congruence-class degenerate representations, with weight
$\vec{r}=\{s+n_1,\dots,s+n_m\}, \ n_1\ge\cdots\ge n_m \in {\bf Z}$
(\ref{chid}). It would be interesting to generalize this construction,
in wiew of describing the experimentally observed filling fractions
$4/11,\ 7/11$, $\ 4/13,\ 8/13$, $9/13,\ 10/17,\dots$, not explained here.
The $\winf$ minimal models can be generalized by considering two (or more)
congruence classes, $\vec{r}=\{s+n_1,\dots,s+n_m; t+k_1, \dots, t+k_l \}$,
with $k_1\ge\dots\ge k_l \in {\bf Z}$ and $c=m+l$.
There are analogies between this mathematical construction
and the Jain hierarchical construction of wave functions, which read
\beq
\Psi_\nu= D^{q/2}\ L^l\ D^{p/2}\ L^m\ {\bf 1}, \qquad\qquad p,q\ {\rm even},
\label{secord}\eeq
to second order of iteration \cite{jain}.
The number of fluids in any $\winf$ congruence
class corresponds to the number of Landau levels in (\ref{secord});
in both constructions, there are two independent elementary anyons,
each one accompanied by neutral excitations.
However, we have not yet proven a complete equivalence of the two
second-order hierarchies: the Jain construction assigns a definite filling
fraction to each wave function (\ref{secord}), while we have a large
modular degeneracy (of the group $SL(2,{\bf Z})$) in the definition
of the physical charge of the two independent anyons, leading to many
values of the filling fraction for each minimal model.
On the contrary, we would like to find {\it more} constraints than in the
Jain construction, because most of its second-order filling fractions are
not observed experimentally.
We guess that our algebraic construction of $\winf$ minimal model
Hilbert spaces should be supplemented by the construction of other physical
quantities, like the partition function \cite{ciz}, which could impose further
conditions on the physical theories.

\bigskip

\bigskip

\bigskip

\noindent{\bf Acknowledgements}

We would like to thank Prof. Victor Kac for triggering our attention to
$\winf$ minimal models. We also enjoyed conversations with Luis Alvarez-Gaum\'e
and Martin Greiter. This work was supported in part by the CERN Theory
Division,
the MIT Center for Theoretical Physics and the European Community program
``Human Capital and Mobility''.

%\vfill\eject
%
%
\def\NP{{\it Nucl. Phys.\ }}
\def\PRL{{\it Phys. Rev. Lett.\ }}
\def\PL{{\it Phys. Lett.\ }}
\def\PR{{\it Phys. Rev.\ }}
\def\IJMP{{\it Int. J. Mod. Phys.\ }}
\def\MPL{{\it Mod. Phys. Lett.\ }}

\end{document}